\documentclass[fleqn,usenatbib]{mnras}
\usepackage{newtxtext,newtxmath}
\usepackage{newtxtext,newtxmath}

\usepackage[T1]{fontenc}
\usepackage{ae,aecompl}

\usepackage{graphicx}	
\usepackage{amsmath}	
\usepackage{hyperref}
\usepackage{natbib}
\usepackage[caption=false]{subfig}
\usepackage{ragged2e}
\usepackage{placeins}
\usepackage{bigints} 
\usepackage{physics}
\usepackage{xcolor}
\usepackage{cleveref}
\usepackage[export]{adjustbox}
\usepackage{ulem}

\newcommand       \ba           {\begin{eqnarray}}
\newcommand       \ea           {\end{eqnarray}}

\title[]{Cold fronts in galaxy clusters I: A case for the large-scale global eigen modes in unmagnetized and weakly magnetized cluster core }

\author[P. P. Choudhury, C. S. Reynolds]
{Prakriti Pal Choudhury$^{1}$ \thanks{Email: prakriti.palchoudhury@physics.ox.ac.uk}, Christopher S. Reynolds$^{2,3}$\\
$^{1}$ Department of Physics, University of Oxford, Parks Rd, Oxford OX13PU, UK\\
$^{2}$ Department of Astronomy, University of Maryland, College Park, MD 20742-2421, USA\\
$^{3}$ Joint Space-Science Institute (JSI), College Park, MD 20742-2421, USA}

\pubyear{2023}

\begin{document}
\label{firstpage}
\pagerange{\pageref{firstpage}--\pageref{lastpage}}
\maketitle
\begin{abstract}
Galaxy clusters show large-scale azimuthal X-ray surface brightness fluctuations known as cold fronts. Cold fronts are argued to originate due to sloshing driven by sub-halo passage at close proximity to the cluster center. While this causes large-scale perturbations, the physical mechanisms that can sustain spiral density structures are not clear. In this work, we explore whether long wavelength thermal instability is an explanation for cold front formation in a cluster core which is perturbed by sub-halos or AGN activity. Using global linear perturbation analysis, we show that unstable internal gravity waves form large-scale three-dimensional spirals, akin to observed cold fronts. We explore if the presence of magnetic field (along spherical $\hat{\phi}$) may support such structures (by suppressing small scale Kelvin-Helmholtz modes) or disrupt them (by promoting additional thermal instability). We find that latter happens at shorter wavelengths and above characteristic Brunt V\"ais\"al\"a frequency ($>N_{\rm BV}$). Our work implies that large-scale spirals are sustained over a long timescale ($>N^{-1}_{\rm BV}$) even in presence of aligned magnetic fields that is otherwise supportive against mixing at the interface. Secondly, short-wavelength (but relatively longer along the field) unstable compressive modes may form within or in the vicinity of such spirals. The instability is an overstable slow wave, and grows in 2D at timescales $\gtrsim 2-3$ times longer than the spiral growth timescale (via thermal instability). Thus this instability cannot destroy the large scale coherence.

\end{abstract}

\begin{keywords}
(magnetohydrodynamics) MHD - galaxies: clusters: intracluster medium 
\end{keywords}



\section{Introduction}
\label{sec:intro}
High resolution X-ray imaging of the intracluster medium of many galaxy clusters show sharp discontinuities in surface brightness, which translate to discontinuities in density (jumps of $\lesssim 100\%$ and $\sim 30 \%$ on an average from outward to CF, which implies a fractional density fluctuation $\sim 0.3 $), metallicity, temperature, and milder discontinuities in pressure. These are referred to as cold fronts (CFs; \citealt{MARKEVITCH20071}). Broadly, according to hydrodynamic simulations and X-ray observations, CFs are classified into merger CFs (contact discontinuities between the ICM of the cluster and subcluster, e.g., \citealt{2023ApJ_sarkar}), stream CFs driven by continuous accretion of cold stream from the IGM to halfway into the ICM (e.g., \citealt{2018MNRAS_zinger}), and CFs seen near the core (e.g., \citealt{2020MNRAS_naor}). In the latter case, thermal pressure discontinuities ($\sim 10-20\%$ jumps from CF outward) have been discussed using deprojected temperature and density in the vicinity of CF (first discussed in \citealt{PhysRevLett_reiss} and then \citealt{2020MNRAS_naor}). These works also discuss the prospect of non-thermal pressure support indicating moderate magnetic field for such a medium. Note that the observational literature on CF preceding the above mentioned works, conclude CFs to be approximately isobaric. While circumstantial evidence of magnetic field already appeared from the width of CF below Coulumb mean free path (e.g., \citealt{2001ApJ_vikhlinin}; also see \citealt{2016JPlPh_zuhone} for CFs as probes of plasma physics), the argument on thermal pressure jump substantiates magnetic field and its coherence aligned with discontinuity. The density contrast is always in the sense that the high density zone sits deeper in the gravitational potential well and, hence, these are considered stable against Rayleigh-Taylor instabilities. The origin of these cooling spirals has been mostly described by interaction of off-center fly-by satellite halos leading to gas sloshing (\citealt{2007ApJ_dupke, 2006ApJ_ascasibar, 2011ApJ_zuhone, 2013ApJ_zuhone}). These works gradually build the picture of CFs with velocity shear at the interface, the development of sufficient magnetic field amplitude to mitigate Kelvin-Helmholtz instabilities across the CFs, and sustenance of metallicity and X-ray brightness discontinuity (\citealt{2011MNRAS_roediger}). Other scenarios of formation include shear driven centripetal acceleration (\citealt{2010ApJ_keshet}). 


Thermal instability (TI) on a background thermal equilibrium has been discussed widely in the context of interstellar medium (\citealt{1977ApJ_mckee}), solar prominences (\citealt{2012ApJ_xia, antolin2022}), accretion disks (\citealt{2000ApJ_nayakshin}), and in the intracluster/circumgalactic medium (\citealt{ 2012MNRAS_mccourt, 2012MNRAS_sharma, 2012ApJKunz, 2021MNRAS_esmerian, donahue2022baryon, 2022MNRAS_mohapatra}) to describe the multiphase nature of the gaseous medium. The latter is the most relevant context for CFs. The idea of generating a radiative cooling driven dense phase from a hot diffuse medium, which is sufficiently heated intermittently, is compelling. The problem can be easily abstracted to instabilities and oscillations about an equilibrium if cooling and heating roughly balance one another (although that is not a necessary condition for thermal instability, see \citealt{1986ApJ_balbus}; such an assumption makes the problem analytical to some extent). Even solar prominence has been explained using models of TI (\citealt{2012ApJ_xia, antolin2022}). There are also TI models for dynamical systems like outflows from accretion disk scales (e.g., \citealt{2022ApJ_waters}), outflows at halo scales (e.g., \citealt{2022ApJ_huang}), and cooling flows in clusters (e.g., \citealt{1986MNRAS_nulsen}). On the other hand, TI has also been discussed at cosmological scales (e.g., \citealt{mandelker2021thermal}). Given the widespread application, it is lucrative to explore if TI can also produce CF geometry (spirals and/or long wavelength in spherical $\hat{\phi}$ direction when line-of-sight is along $\hat{z}$ without loss of generality).

In galaxy cluster cores, isobaric local thermal instability is often evoked to explain H$\alpha$ filaments seen in observations (\citealt{2001MNRAS_fabian, 2016MNRAS_fabian}). The current paradigm is that a cooler, dense phase condenses from the hot intracluster medium (ICM) due to local TI, with a morphology dictated by local magnetic field structure (\citealt{2018MNRAS_ji,2024MNRAS_das}). Azimuthal large-scale density contrasts, if present, are expected to be global modes instead of local TI. CFs, described by global thermal instability, must also produce metallicity jumps (\citealt{2010MNRAS_simionescu}). More recently, it is suggested that metals can be advected by spirals from inside and outside the arms (\citealt{2020ApJ_naor}). Notably, advection of metals, and a resultant staircase structure of metallicity profile (due to discontinuities at the location of the mode) have been discussed in isochoric/isobaric thermal instability (\citealt{2021MNRAS_das}). Thus, global TI may exhibit the necessary properties of CFs.

In this work, we show that the large scale spirals are formed in stratified cluster cores due to thermally unstable buoyancy waves. Further we consider the effect of weak magnetic  fields (high plasma $\beta$, the ratio of thermal to magnetic pressure) since $\beta$ in cluster deduced from Faraday rotation is large (e.g., \citealt{2016A&A_bohringer}). Magnetic fields are expected to be aligned with cold front so that thermal conduction is reduced in the direction perpendicular to the field to preserve the temperature jump seen in observations (\citealt{2003ApJmarkevitch, 2013ApJ_zuhone} \footnote{The exact details of the velocity space anisotropy driven plasma phenomena is missing and is non-trivial to explore in the global cluster scale simulations. The collective effect of scattering between unstable electromagnetic waves and heat carrying electrons and the Coulumb collisions is not well explored, but for example see \citealt{2021ApJ_drake} for a discussion on whistler instability and heat flux.}). 
There is no hint apriori that such a coherent weak field is conducive for or preventive against thermal instability. We analyse if the latter scenario is likely. 

In the absence of magnetic field, the compressive modes are expected to propagate the fastest and typically these do not contribute to the thermal instability in hot galaxy cluster cores. This is easy to see from the instability threshold in which the isentropic modes never satisfy the threshold for linear instability (\citealt{field1965}). On the other hand, the internal gravity waves that are easily confined within the core contribute efficiently to non-linear random shear motions which provide the seed density contrasts to isobaric thermal instability. The threshold for latter is extremely easy to reach in cores. Linearly, these waves are unstable in a characteristic thermal instability (cooling) timescale ($\sim t^{-1}_{\rm TI} \sim t^{-1}_{\rm cool}$). As a result, often the buoyancy oscillations and thermal instability can be thought to be working in conjunction towards growth as well as saturation of condensed gas. The small-scale (relative to core size) density perturbations grow, saturate, and drop out of the hot phase into the gravitational potential well fast. With idealized magnetic field topology, local periodic box simulations of stratified gas shows elevated rates of growth in condensation (e. g.,  \citealt{2018MNRAS_ji}) due to magnetic tension preventing buoyancy oscillations. For an initially oscillating and rapidly cooling clump, magnetic tension reduces the oscillation and supports further cooling {\it in-situ}. 

The buoyancy oscillations are also critically dependent on the characteristic length scales perpendicular to the direction of gravity. For large perpendicular length scales, in principle, the saturation of thermal instability can be delayed since the buoyancy timescale increases (frequency $\sim \frac{k_{\perp} N_{\rm BV}}{k}$, where $N_{\rm BV}$ is a characteristic Brunt-V\"ais\"al\"a frequency for buoyancy oscillations). Since the small-scale growth cannot completely destroy such larger structures (linear growth rate is scale independent), this is a viable mechanism to sustain long filamentary structures of $\sim 100$s kpc. In practice, Kelvin-Helmholtz instability takes over fast. But this can be prevented with a magnetic field aligned with such a filament of thermally unstable gas. This is expected to work perfectly if the alignment and gravity are perpendicular. The question that we address here is: can such a mode describe the cold fronts in clusters? In space plasma (e.g., earth's magnetosphere and the interstellar medium) such a problem has been explored earlier to understand hydromagnetic instability (e.g., \citealt{1959JGR_gold, 1966ApJ_parker}), albeit the field strengths are expected to be stronger than in cluster cores. For a global model of cluster, such an azimuthal field is not entirely justified unless the field is dragged and wrapped around by multiple minor mergers or AGN jets produce such coherent fields via Biermann battery. For the current purpose, we use an idealized azimuthal magnetic field.

The two main issues following the above discussions are (i) can TI be applied to produce azimuthally extended linear density contrasts at moderate amplitude ($\delta \rho/\rho_0 \lesssim 0.1-0.3$)? (ii) Does the presence of weak magnetic fields trigger any other instability to support/destroy such a density contrast if produced? In this first paper, we discuss global linear eigen spectra for thermal instability which produce perfect spirals in hydrodynamic case. We further discuss ideal MHD eigen spectra to assess if that introduces any invasive new instability. In a forthcoming paper, we will discuss a suite of non-linear MHD simulations focussing on the large-scale azimuthal density contrasts in galaxy cluster cores.  

We organize this paper as follows. In section \ref{sec: physset} have two parts, section \ref{sec:globalsetup} presents the method and extension to solve for linear global overstable TI modes with and without magnetic field, and section \ref{sec:analyticsetup} discusses the physical interpretation of magnetized modes.  In section \ref{sec: results} we discuss the results with and without magnetic field in the global atmosphere, and the interpretation using local linear analysis that is introduced in section \ref{sec:analyticsetup} first.
In Section \ref{sec:discnconc}, we discuss and conclude.

\section{Physical models and Methods}
\label{sec: physset}
We explore long wavelength modes by a global linear mode analysis for an intracluster medium with gradients in density, temperature, and pressure. In order to support and interpret the MHD results we also perform a local linear analysis and identify the relevant mode at play. 
In what follows, we first describe the full magnetohydrodynamic equations together with our assumptions about radiative cooling and heating. We then describe the method of global mode analysis in section \ref{sec:globalsetup}, and the local analysis in section \ref{sec:analyticsetup}. 

We consider the following ideal MHD conservation equations for the ICM: 
\ba
\label{eq:eb1}
\frac{D \rho}{Dt} &=& - \rho {\bf \nabla \cdot \bf{v}}, \\
\label{eq:eb2}
\frac{D \bf{v}}{Dt} = -\frac{1}{\rho} {\bf \nabla} p &-& g \bf{\hat{r}} + \frac{1}{\rho} ({\bf \nabla}\times {\bf B}) \times {\bf B}, \\
\label{eq:eb3}
\frac{p}{(\gamma-1)} \frac{D}{Dt} \left [ \ln \left ( \frac{p}{\rho^\gamma} \right ) \right ] &=& -q^-(n,T) + q^+(r,t), \\
\label{eq:eb4}
\frac{D \bf{B}}{Dt} &=& ({\bf{B}} \cdot {\bf \nabla}) {\bf v} - {\bf{B}} ({\bf \nabla} \cdot {\bf v} )
\ea
where $D/Dt$ is the Lagrangian derivative, and $\rho$, $\bf{v}$ and $p$ are mass density, velocity and 
pressure.  Radiative cooling is described by the term $q^-(n,T) \equiv n_e n_i \Lambda(T)$ is the radiative cooling
(where $n_e \equiv \rho/[\mu_e m_p]$ and $n_i \equiv \rho/[\mu_i m_p] $ are electron and ion number densities, respectively; $\mu_e=1.17$, $\mu_i=1.32$, 
and $m_p$ is proton mass, $\Lambda(T)$ is the temperature-dependent cooling function).  We include a heating term that, by assumption, balances the radiative cooling in a shell-averaged sense, $q^+(r,t) \equiv \langle q^- \rangle$, and represents AGN feedback via local turbulent heating (also see section 5.4 in \citealt{2012MNRAS_mccourt} for an analysis of this assumption).

Assuming the magnetic field is sufficiently weak that the gas pressure dominates, we assume that background hydrostatic equilibrium implies $dp_0/dr = -\rho_0 g$, where a subscript `0' refers to equilibrium quantities and acceleration due to gravity
$g \equiv d\Phi/dr$ ($\Phi$ is the fixed NFW gravitational potential). The details of the atmosphere and the physical parameters are mentioned in \citealt{2016MNRAS_choudhury}. These equations are closed by the divegence free condition, ${\bf \nabla} \cdot {\bf{B}} = 0$.

\subsection{Global perturbations}
\label{sec:globalsetup}
We first write the linearised equations. The perturbations, the background magnetic field orientation and the specific considerations are described below the following linearized equations. 
\ba 
\label{eq:ee1}
\sigma F_\rho &=& - \frac{1}{r^2} \frac{d}{dr} (r^2 \rho_0 F_r) + \underbrace{l(l+1) \frac{\rho_0 F_\theta}{r}}^\text{},\\
\nonumber
&&\Big[-[m^2 - l(l+1)] \frac{\rho_0 F_\theta}{r} + \frac{m^2 \rho_0 F_\phi}{r} \Big]\\
\label{eq:ee2}
\sigma F_r &=&   -\frac{1}{\rho_0} \frac{d}{dr} \left [ p_0 \left ( \frac{F_s}{s_0} + \gamma \frac{F_\rho}{\rho_0} \right ) \right ]-\frac{gF_\rho}{\rho_0}, \\
\nonumber
&&\Big[- \frac{1}{\rho_0} \frac{\partial}{\partial r} \Big( \frac{B_0}{r}\frac{\partial (r F_{\rm A})}{\partial r} \Big) + \Big(m^2F_{\rm A} - \frac{2\partial (r F_{\rm A})}{\partial r} \Big)\frac{B_0}{\rho_0 r^2} \Big]\\
\label{eq:ee3}
\sigma F_\theta &=& -\frac{p_0}{r \rho_0} \left ( \frac{F_s}{s_0} + \gamma \frac{F_\rho}{\rho_0} \right ), \\
\nonumber
&&\Big[ - \frac{1}{r \rho_0} \Big( \frac{B_0}{r}\frac{\partial (r F_{\rm A})}{\partial r} \Big]\\
\label{eq:ee4}
\sigma F_\phi &=& -\frac{p_0}{r \rho_0} \left ( \frac{F_s}{s_0} + \gamma \frac{F_\rho}{\rho_0} \right )\\
\nonumber
&&\Big[ - \frac{1}{r \rho_0} \Big( \frac{B_0}{r}\frac{\partial (r F_{\rm A})}{\partial r} \Big) + \frac{B_0}{\rho_0 r} \frac{\partial (F_{\rm A})}{\partial r} - \frac{F_{\rm A}}{\rho_0 r} \frac{\partial B_0}{\partial r}\Big]\\
\label{eq:ee5}
\sigma F_s &=& -\frac{\gamma s_0 N_{\rm BV}^2}{g} F_r \\
\nonumber
&-& \frac{s_0}{t_{\rm cool 0}} \left [ 2 \frac{F_\rho}{\rho_0} + \frac{d\ln \Lambda}{d\ln T} \left( \frac{F_s}{s_0} + (\gamma-1) \frac{F_\rho}{\rho_0} \right) \right]\\
\label{eq:ee6}
 \sigma F_{\rm A} &=& - F_{\rm r} B_0
\ea 

In this section, we look for the three-dimensional perturbations in a global spherical cluster atmosphere but on the plane of $\theta =\pi/2$ (easily tractable and without loss of generality we basically take a plane with magnetic field and gravity; the equations are first linearized and then we use $\theta=\pi/2$).\footnote{In the most general formulation of our global linearized equations, for any $\theta$, trigonometric functions of $\theta$ appear. The same exercise can be trivially extended for various values of $\theta$. It will reveal the entire 3D structure of the instability. Further, due to this reason, the instability disappears when we use poloidal field. The presence of background field on the plane ($\theta=\frac{\pi}{2}$ condition) is essential to assess the instability that we discuss in ideal MHD.} Note that in the hydrodynamic case, this assumption of $\theta = \pi/2$ is not a necessity in this formalism. The background quantities describing the equilibrium are radial functions (see \citealt{2016MNRAS_choudhury}) and 
the perturbations, in general, depend on all coordinates and time as $e^{\sigma t} \delta_{\rho, p, s, v} (r, \theta, \phi)$.
Our global linear stability analysis is solved as a linear eigenvalue problem in radius. The radial component of all perturbed quantities (density, 
velocity, etc.) are expanded in a Chebyshev polynomial basis (Pseudospectral method; see \citealt{boyd2001chebyshev}), and the matrix equation for eigenvalues and eigenfunctions is solved numerically on a Gauss-Lobatto (GL) grid of size $n$ (sets the number of terms in the basis as well). The GL grid is a mapping of the original grid $[r_{\rm in}, r_{\rm out}]$ to a new variable $-1<=\zeta<=1$ and all boundary conditions are put at $\zeta=[1,-1]$. \footnote{This method is associated with application of Gaussian quadrature (integral) in spectral methods to solve eigen-problem and Gauss-Lobatto grid provides efficient quadrature rules. In this method, it is important to assess convergence; in other words, for sufficiently high values of $n$, only the physical eigenvalues for every $n$ we try, must match at high precision.} In addition to the boundary conditions used in \citealt{2016MNRAS_choudhury} (section 3, last paragraph), we use a boundary condition for $F_{\rm A}$ as described later (below).
The perturbed quantities are written in a spherical harmonic basis in the angular direction as follows,
$$
\delta \rho = F_\rho (r) Y_l^m,
\delta p = F_p (r) Y_l^m,
\delta s = F_s (r) Y_l^m,
\delta T=F_T (r) Y_l^m,
$$
$$
\delta v_{r} = F_r (r) Y_l^m,
\delta v_{\theta } = F_\theta (r)  \frac{\partial Y_l^m(\theta,\phi)}{\partial \theta},
\delta v_{\phi } =  \frac{F_\phi (r)}{\sin \theta} \frac{\partial Y_l^m(\theta,\phi)}{\partial \phi},
$$
$$\delta A = F_{\rm A} (r) Y_l^m$$
where $Y_l^m(\theta,\phi)$ are the spherical harmonics of order $(l,m)$ and $F$ carries the radial dependence. Here, $s = \ln \left ( \frac{p}{\rho^\gamma} \right )$ denotes the entropy index governed by Eq. \ref{eq:eb3} and $N_{\rm BV}= \frac{g}{\gamma}\frac{\partial s}{\partial r}$ is the characteristic Brunt-V\"ais\"al\"a frequency for buoyancy oscillations (also mentioned in the introduction). These forms 
are obtained by comparing the $r,~\theta,~\phi$ dependence of various terms in Eqs. \ref{eq:eb1}-\ref{eq:eb4}. In hydrodynamics, if we write the equations corresponding to $F_\phi$ and $F_\theta$, these have identical evolution (related to the pressure gradient term). Hence mathematically, solving for $F_{\theta}$ or $F_{\phi}$ in hydrodynamic case is equivalent (note that there are differences in $\theta$ and $\phi$ dependence for the $\delta v_{\theta}$ and $\delta v_{\phi}$). In magnetohydrodynamics, this degeneracy of the evolution of the radial part of the angular velocity perturbations is broken by the magnetic field direction and we need to consider both equations in the dynamics. $\delta A$ is the perturbation of the magnitude of background magnetic vector potential ($\hat{\theta}$) that we describe below. 

 We first perturb the equilibrium and then solve for the perturbed quantities only on the central plane ($\theta = \frac{\pi}{2}$). In eq \ref{eq:ee1}, if $F_{\theta} = F_{\phi}$ (the two non-radial directions are indistinguishable at any given radial point in the unmagnetized case), then we can directly use 
 \ba
 \frac{1}{\sin \theta} \frac{\partial}{\partial \theta} \Big( \sin \theta \frac{\partial Y^m_l}{\partial \theta} \Big) + \frac{1}{\sin^2 \theta}\frac{\partial^2 Y^m_l}{\partial \phi^2} = - l(l+1)Y^m_l 
 \ea
 but if $F_{\theta} \neq F_{\phi}$ we can instead utilize another property of spherical harmonics and write, 
 \ba
 \frac{1}{\sin \theta} \frac{\partial}{\partial \theta} \Big( \sin \theta \frac{\partial Y^m_l}{\partial \theta} \Big) = \Big( \frac{m^2}{\sin^2 \theta} - l(l+1) \Big) Y^m_l
 \ea
 The latter is used in the first equation. In what follows, we describe the terms associated with ideal MHD.
 Note that we perturb the magnetic vector potential of the guide field in this analysis above to assure divergence-free condition (where ${\bf \nabla} \times (A \hat{\theta}) = {\bf B}, A=A_0(r) + \delta A$). 
Also, eqns.~\ref{eq:ee1}-\ref{eq:ee6} assume the background magnetic field is in $\hat{\phi}$,
\ba
\label{eq:B}
{\bf \nabla} \times {\bf A_0} = {\bf B_0} =  B_0(r) \hat{\phi} 
\ea
where $B_0 (r) = \sqrt{\frac{8\pi}{\beta}}\sqrt{p_0 (r)}$.
Due to the perturbation $\delta A$, there is perturbation in magnetic field along $\hat{r}$ and $\hat{\phi}$, such that $\delta B_{\rm r} = -\frac{1}{r \sin{\theta}} \frac{\partial \delta A}{\partial \phi}$ and $\delta B_{\phi} = \frac{1}{r}\frac{\partial (r \delta A)}{\partial r}$.
The induction equation in terms of vector potential is,
\ba
\label{eq:ind}
\frac{\partial {\bf A}}{\partial t} = {\bf v} \times ({\bf \nabla} \times {\bf A})
\ea
which can be used to track the evolution of $\delta A$. Here we assume Weyl gauge or electric scaler potential to be zero. Note that the right hand side in the above equation is dependent only on the background magnetic field $B_0$ (in ${\bf v} \cross B_0 \hat{\boldsymbol \phi}$) since additional effects will be non-linear (velocities are linear perturbations). Thus in this system, we simply need to track the perturbation $\delta A$ ($\delta A_{\theta}$) in ${\bf A_0}$ (or the equivalent, $\bf{A_{\theta}}$). However, one limitation is that $\delta A_{\rm r}$, which can feed $\delta B_{\theta}$ and $\delta B_{\phi}$ via the derivatives $\frac{1}{r \sin{\theta}} \frac{\partial \delta A_{\rm r}}{\partial \phi}$ and $-\frac{1}{r}\frac{\partial \delta A_{\rm r}}{\partial \theta}$ is not evolved in our system. Firstly, this limitation makes the equations analytically tractable. Secondly, this is not problematic for our exploration of long wavelength modes along $\hat{\phi}$ in the $\theta=\pi/2$ plane. To elucidate the consequences of this limitation and why it is not an issue, we write the induction equation for $\delta A_{\rm r}$ and the complete $\bf {\delta B}$ (including the terms that are ignored in our approach) below:
\ba
\nonumber
\sigma F_{\rm A_{\rm r}} &=&  F_{\theta} B_0\\
\nonumber
{\bf{\delta B}} = - \frac{1}{r \sin \theta}
\frac{\partial \delta A_{\theta}}{\partial \phi} \hat{\bf r} &+& \frac{1}{r \sin \theta}
\frac{\partial \delta A_{\rm r}}{\partial \phi} \hat{\boldsymbol \theta}\\
+ \frac{1}{r} \Big( \frac{\partial}{\partial r}(r \delta A_{\theta})  &-& \frac{\partial \delta A_{\rm r}}{\partial \theta} \Big) \hat{\boldsymbol \phi}
\label{eq:delB}
\ea
Ignoring $\delta A_{\rm r}$ means we assume perturbation in velocity along $\hat{\boldsymbol \theta}$ is much smaller than velocity perturbation along $\hat{\bf r}$ (verified for the relevant global modes we discuss in Figure \ref{fig:app2}). If we pursue long wavelength (along $\hat{\boldsymbol \phi}$) modes in $r-\phi$ plane, in the spherically symmetric background, our limitation of assuming small $\delta A_{\rm r}$ compared to $\delta A_{\theta}$ is not unreasonable. We are choosing modes with velocity perturbations, $F_{\theta}<<F_{\rm r}$ and $\sigma F_{\rm A_{\rm r}} \approx 0$ (selectively capturing motions on the plane of gravity and magnetic field but retaining the parallel and perpendicular directions to the field). The divergence-free condition of $\mathbf{\delta B}$ remains unchanged. We also reduce the equations easily into analytically tractable form with our limitation.\footnote{On the other hand, if we take $\delta A_{\rm r} = F_{\rm A_r} \frac{\partial Y^m_l}{\partial \theta}$ in the same form as $\delta v_{\theta}$, we end up obtaining higher $\theta$ derivatives of spherical harmonics in the associated magnetic pressure perturbations and tension terms. That makes the linear problem far more complex.} In eq \ref{eq:delB}, we thus include only the terms associated with $\delta A_{\theta}$ (alternately $\delta A$, the only component considered in magnetic potential perturbation).

In eq \ref{eq:eb2}, we expand the Lorentz force term into the gradient of magnetic pressure and magnetic tension as following: $({\bf \nabla}\times {\bf B}) \times {\bf B} = \Big( -{\bf \nabla} (B^2/2)+ ({\bf B}\cdot{\bf \nabla}) {\bf B} \Big)$. Here, magnetic pressure ($p_{\rm B} = B^2/2$) and magnetic tension [$=({\bf B}\cdot{\bf \nabla}) {\bf B}$] are linearised as $-{\bf \nabla}({\bf B_0}\cdot{\delta {\bf B}}) + ({\bf B_0 \cdot \nabla}) \delta {\bf B} + ({\bf \delta B \cdot \nabla}) {\bf B_0}$. The perturbations $\delta B_{\rm r}$ and $\delta B_{\phi}$ fed by $\delta A$ ($\delta A_{\theta}$ in case of field in $\hat{\phi}$ direction) can be expressed in terms of $F_{\rm A}$ (the radial part of the perturbation $\delta A$). Hence the perturbation associated with magnetic pressure is $\delta p_{\rm B} =  B_0 \delta B_{\phi} = \frac{B_0}{r}\frac{\partial (r \delta A)}{\partial r}$. If we perturb the tension term, we get the following vector (the components of which can again be expressed in terms of derivatives of $F_{\rm A}$):
\ba
\nonumber
\Big( \frac{B_{0}}{r \sin{\theta}} \frac{\partial \delta B_{\rm r}}{\partial \phi} - \frac{2 B_{0} \delta B_{\phi}}{r}\Big) \hat{\bf r} &-& \Big(\frac{2 B_{0} \delta B_{\phi} \cot{\theta}}{r} \Big) \hat{\boldsymbol \theta} \\
\nonumber
+ \Big(\frac{B_{0}}{r \sin{\theta}} \frac{\partial(\delta B_{\phi})}{\partial \phi} &+& \frac{B_{0} \delta B_{\rm r}}{r} + \delta B_{\rm r} \frac{\partial B_{0}}{\partial r} \Big) \hat{\boldsymbol \phi}
\ea
The final expressions in which all MHD terms are written in terms of $F_{\rm A}$ are written in eqs \ref{eq:ee1} - \ref{eq:ee6}. 

Lastly the boundary condition for $F_{\rm A}$ is $\frac{ dF_{\rm A}}{d\zeta} = 0$ at inner radial point where $\zeta$ denotes the mapped grid from the physical grid. The boundary conditions may not be unique to obtain same solutions (see last paragraph of section 3 in \citealt{2016MNRAS_choudhury}). Moreover, $F_{\rm r} = F_{\phi} = F_{\rho}=0$ at the outer radial point, $\frac{d F_{\theta}}{d\zeta}=0, F_{\rm s}=0$ at the inner radial location.

\subsubsection{Comparison with previous ideas on spiral flows in clusters}
In earlier works, a comprehensive discussion on the co-evolution of magnetic field, thermally balanced cool-cores (CCs) and thermal instability is missing. Our analysis is a step ahead from that point. Although sloshing simulations successfully produce the observed spirals, these are often argued to be transient features due to CF propagation, along with disruption by thermal instability and turbulence/mixing (the destructive influence of turbulence induced by the merger itself is tested recently in \citealt{2024ApJ_bellomi} in the absence of radiative cooling and energy sources). In the presence of radiative cooling, without adequate energy injection, the CFs can lose hydrostatic balance and/or cool vigorously, ultimately contributing to the cooling flow problem (\citealt{fabian1994cooling}). The latter scenario does not occur in previous substructure simulation (\citealt{2010ApJ_zuhone}) possibly due to the addition of entropy by the substructure passage (and/or entropy variation due to resolution constraints as mentioned in section 2.2), and the ``switch-on" of radiative cooling after the injection of entropy (Fig. 8 and last column of Table 2 in \citealt{2010ApJ_zuhone}). The cooling catastrophe occurs in the same suite of simulations when viscosity and magnetic field are included (\citealt{2011ApJ_zuhone}). It is not clear whether small-scale thermal instabilities appear in the core of these simulations and/or if such instabilities are enhanced in the presence of a magnetic field and eventually contribute to a cooling catastrophe. In order to assess CF evolution, sufficient understanding of the multi-scale instabilities in presence of radiative cooling, AGN-motivated heating and magnetic field is necessary. 

In the absence of magnetic field, \citealt{2012ApJ_keshet} explore the consequences of an isobaric spiral large scale rotational (slowly) flow in CCs and argue on the mediatory role of CFs to quench cooling flows by mixing X-ray deficient radio bubbles entrained in the spirals (although no source of heating is considered in the actual calculation unlike our analysis). They derive a key physical insight that the radius at which the spiral forms must be smaller than the radius of curvature corresponding to extension in $z-r$ plane (or equivalently, $\mathbf{\hat{\theta}}$), and thus more cylindrical spirals. This is supported by the argument that flow planes of spirals cannot intersect or interact. On the contrary, in the MHD context, we consider global flow velocities along $\mathbf{\hat{\theta}}$ to be much smaller (but not identically zero) than flow velocity along $\mathbf{\hat{r}}$ in the equatorial plane and we take no assumption for flow along $\mathbf{\hat{\phi}}$. This is appropriate for the magnetic field geometry (along $\mathbf{\hat{\phi}}$) and the stratification direction (along $\mathbf{\hat{r}}$). As we use spherical harmonics for basis, we also do not have cylindrical geometry of the CFs and as mentioned in section \ref{sec: results}, the perpendicular scale (along $\mathbf{\hat{\theta}}$) is fixed by moderately large $l$ (shorter length). Further, we do not necessarily assume isobaric modes. However, we find that (i) large spirals are isobaric with and without magnetic field, (ii) for too large $l$ in MHD case (not in hydrodynamics), spiral instabilities disappear. In the hydrodynamic case, we also find that for the spiral instabilities of a given $l$ and varying $m$, the observer moving towards an edge-on view will see more concentric semi-circular and vertical structures. 

More recently, \citealt{2024MNRAS_roediger} discuss a toy model as a series of classical oscillators with characteristic buoyancy wave frequency at any given radius. Such wave-like propagating CFs (similar to what we discuss in detail here) have been distinguished from sub-structure driven CFs since in latter, material in the ICM and the sub-structure can contribute or modify the CF features. The argument put forward by this work regarding the sustenance of a wave-like CF is that by the time Kelvin-Helmholtz rolls grow significantly (a few growth times), the wavefront propagates the distance equivalent to its width. We disagree that this should be the sufficient condition for the CF to survive since the moving interface of the ICM and CF should be still rolled up and become non-linear by a few growth times (that is the generic characteristic of any instability). The key reason to claim that the hydrodynamic buoyancy waves in our work should satisfy the conditions of survival is that the dense regions are simultaneously growing due to thermal instability within a comparable timescale to that of propagation. Moreover, we further include aligned magnetic field (that can support against mixing) to assess if (i) spirals still form, and (ii) there are any additional instabilities. We find that if spirals form, these are still growing at radiative/buoyancy timescales and small scale modes primarily will make a growing CF wiggle and relax periodically along the field. Strictly speaking, such wiggles are also growing (as we discuss in section \ref{sec: results}) but at least an order of magnitude below the growth rate of the CF in the global case. The only tentative channel of CF destruction in our proposed scenario might be from equivalent local thermal instabilities which typically grow at the same rate as that of global growth, independent of the length scales.

\subsection{The physical nature of global modes}
\label{sec:analyticsetup}

\begin{figure*}
    \includegraphics[width=18cm]{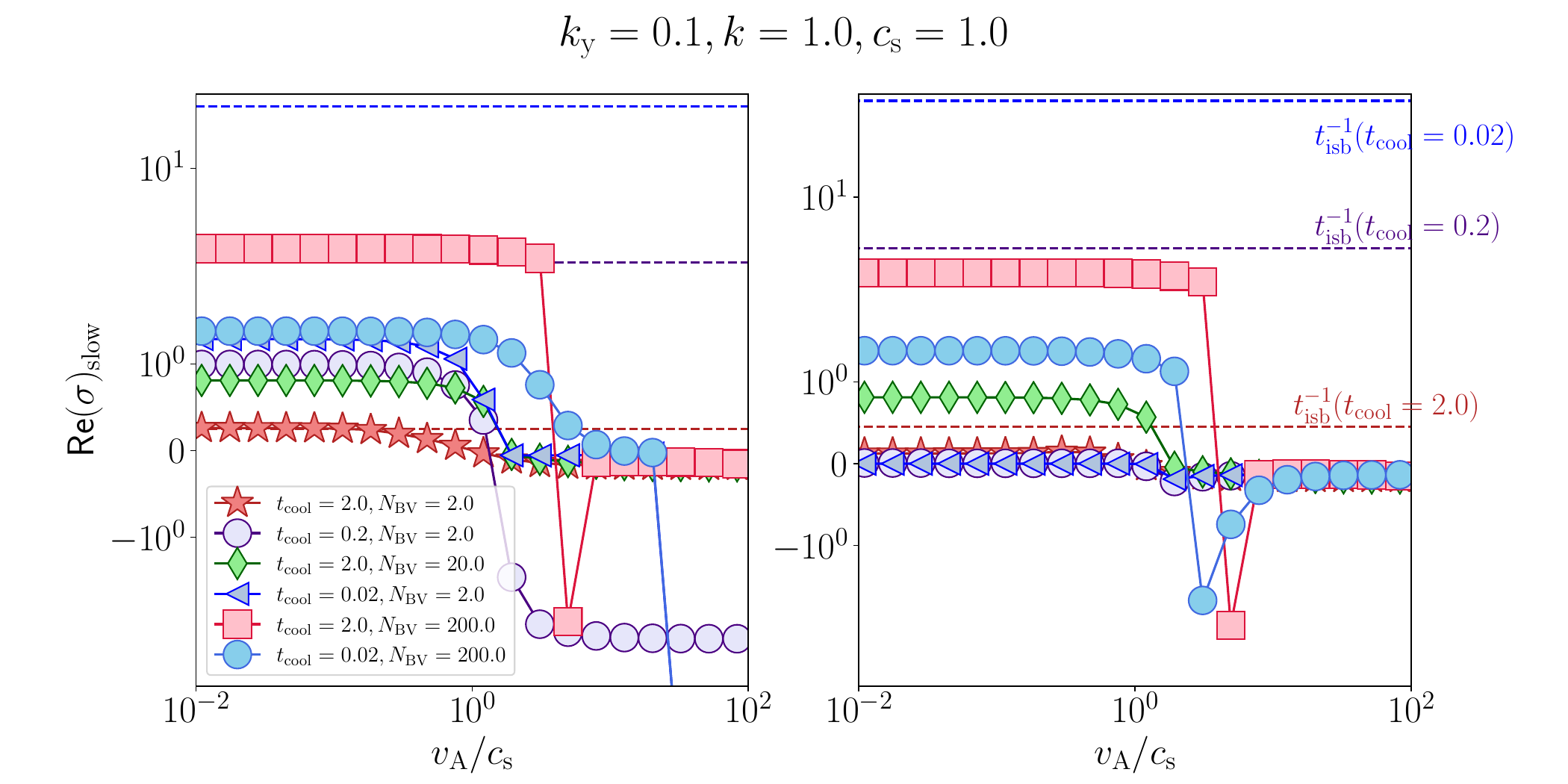}
    \centering
    \caption{The local growth rates of the slow waves versus $v_{\rm A}/c_{\rm s}$ as solved using eq. \ref{eq: disp2} with arbitrary physical parameters like $t_{\rm cool}, N_{\rm BV}$, etc (Figure \ref{fig:localmodes} shows with realistic local physical parameters for a cluster core). The purpose is to demonstrate the growth rates relative to $t^{-1}_{\rm isc}$ (in left panel) and $t^{-1}_{\rm isb}$ (in right panel; the three horizontal dashed lines in red, purple, and blue to be compared to red star, purple circle and blue triangle; $t^{-1}_{\rm isb}/t^{-1}_{\rm isc}$ is $\frac{2-\Lambda_{\rm T}}{\gamma \Lambda_{\rm T}}= \frac{9}{5}$). $k_{\rm y}$ is chosen to be small to allow for longer modes along the local magnetic field. If $k_{\rm y}\rightarrow k$, the growth rates are higher. If $k_{\rm y}\rightarrow 0$, left mode is unstable at smaller rates and the right transitions to stability. At $v_{\rm A}/c_{\rm s} >> 1$ no growth happens.}
    \label{fig:growthlocalgen}
\end{figure*}
In this section we present a physical interpretation of the global modes (discussed in next section) in the light of a local analysis and previous literature.
We take a simplified local model to compare with and predict the physical origin of the global modes. The background (not the perturbations) thermodynamic variables like density, pressure, temperature, and magnetic field are imposed to be locally constants in this analysis. We will use the spatial location of the global instability and use the respective $t_{\rm cool}$, $N_{\rm BV}$, and similar parameters in this local analysis to understand which mode appears globally. We now describe the formulation of this local analysis.

\ba
\nonumber
\sigma \frac{\delta \rho}{\rho_0} &=& - i \mathbf{k}\cdot \mathbf{v} \\
\nonumber
\sigma \mathbf{v} &=& -iv^2_{\rm t} \mathbf{k}\frac{\delta p}{p_0} - \frac{i}{\rho_0}\mathbf{k}(\mathbf{B_0} \cdot (\mathbf{k} \cross \delta {\mathbf A})) \\
\nonumber
&+& \frac{i}{\rho_0}(\mathbf{B_0} \cdot \mathbf{k}) (\mathbf{k} \cross \delta \mathbf{A})) -g\hat{x} \frac{\delta \rho}{\rho_0}\\
\nonumber 
(\sigma + \frac{1}{t_{\rm isc}})\frac{\delta p}{p_0} &=& - \frac{\gamma N_{\rm BV}^2}{g} \hat{g} \cdot \mathbf{v} + \gamma (\sigma - \frac{1}{t_{\rm isb}})\frac{\delta \rho}{\rho_0}\\
\nonumber
\sigma \delta A &=& v_{\rm x} B_0
\ea
where the standard isobaric and isochoric growth/decay timescales are $t_{\rm isb} = \frac{\gamma t_{\rm cool}}{2-\Lambda_T}$, $t_{\rm isc}= \frac{t_{\rm cool}}{\Lambda_T}$, $\gamma  v^2_{\rm t} = c^2_{\rm s}$ and $\Lambda_T = d\ln \Lambda/d \ln T$. Here, $N_{\rm BV}$ is the Brunt-V\"ais\"al\"a frequency (which is the largest frequency of stable buoyant oscillations), and $B_0$ is the background magnetic field with $\beta = \frac{8\pi p_{\rm th}}{B^2_0} >> 1$ so that the background equilibrium is closely approximated by the hydrodynamic case. Note that the second and the third terms in the velocity/momentum equation stand for the magnetic pressure [$-\frac{i}{\rho_0}\mathbf{k}(\mathbf{B_0} \cdot \delta {\mathbf B})$] and magnetic tension [$\frac{i}{\rho_0}(\mathbf{B_0} \cdot \mathbf{k})\delta \mathbf{B})$]. Before delving into the system, we briefly discuss previous exploration of similar magnetospheres.

\citealt{ferriere2001interchange} (Fig.~2) shows that both slow wave and Alfv\'en wave become overstable in presence of magnetic field due to buoyancy driven instabilities which have diverse names in the literature. Mathematically, modes associated with strictly transverse motions are called type I and those associated with strictly longitudinal motions are called  type II interchange modes. The first case physically means a whole flux tube is displaced (or interchanged) and is often considered a generalization of hydrodynamic gravity mode; while in the second case the field line ripples and generates gravity and pressure-driven forces despite line aligned motions of the medium itself. However, motions in different directions are coupled and no such strictly type I or II exist in real magnetospheres. Doing away with the formal definitions, \citealt{1999JGR_ferriere} analyse a system with horizontal field (including or excluding the field curvature $\mathbf{\hat{b}}\cdot \boldsymbol {\nabla} \mathbf{\hat{b}}$), and vertical stratification to understand unstable modes driven by buoyancy in presence/absence of rotation. While we do not consider rotation in the medium, thermal instability (uneven local radiative cooling) can give rise to the coupling terms that support growth of instability, similar to what rotation does.

Since motions in parallel and perpendicular directions (to background field) are used in literature to categorize this class of buoyancy instabilities, we consider the momentum equations closely. The projections of vector momentum equation along the (i) propagation direction, (ii) direction of stratification and gravity, and (iii) background magnetic field direction, reveal the driving forces as following:
\ba
(\sigma^2 - i \mathbf{k}\cdot \mathbf{g}) \frac{\delta \rho}{\rho_0} &=& - k^2 \delta p_{\rm tot} + \mathbf{k} \cdot \delta \mathbf{F}_{\rm T}\\
\nonumber
\frac{\delta \rho}{\rho_0} &=& \Bigg[ \frac{(\sigma + t^{-1}_{\rm isc})\frac{\delta p}{p_0}}{\gamma (\sigma - t^{-1}_{\rm isb})} + \frac{\mathbf{v}\cdot \boldsymbol{\nabla} s_0}{\gamma (\sigma - t^{-1}_{\rm isb})}\Bigg]\\
\nonumber
\Big(\sigma + \frac{N^2_{\rm BV}}{\sigma - t^{-1}_{\rm isb}} \Big) \mathbf{v}\cdot \boldsymbol{\nabla} s_0 &=& -i (\mathbf{k}\cdot \boldsymbol{\nabla} s_0) \delta p_{\rm tot}\\ - \frac{N^2_{\rm BV}(\sigma + t^{-1}_{\rm isc})\frac{\delta p}{p_0}}{(\sigma - t^{-1}_{\rm isb})} &+& i \delta \mathbf{F}_{\rm T} \cdot \boldsymbol{\nabla} s_0\\
\sigma \mathbf{v}\cdot \mathbf{B}_0 &=& -i \mathbf{k} \cdot \mathbf{B}_0 \delta p_{\rm tot} + i \delta \mathbf{F}_{\rm T}\cdot\mathbf{B}_0 
\label{eq:momdot}
\ea
where $s_0 = \ln \frac{p_0}{\rho^{\gamma}_0}$, perturbations in magnetic pressure and tension are included as $\delta p_{\rm tot} = \gamma^{-1} c^2_{\rm s} \frac{\delta p}{p_0} + \frac{\mathbf{B_0}\cdot \mathbf{\delta B}}{\rho_0}$ and $\mathbf{\delta F}_{\rm T}=\frac{(\mathbf{B_0}\cdot \mathbf{k})\mathbf{\delta B}}{\rho_0}$. The first equation emphasizes that the velocity along the direction of wave propagation is sourced by change in density contrast (thermal instability) which leads to a magnetic tension as well. Along the direction of stratification, an imbalance between buoyancy, pressure and tension can produce motions. Along the background field (last equation), there is less chance of motions without the intervention of radiative cooling (via the thermal pressure term). In fact, for a pure isobaric case, only motions along gravity are expected leading to a generalized buoyancy overstability (what should also be categorized in literature as type I). We can see this by simply considering $\delta p_{\rm tot}\approx 0, \delta p/p_0 \approx 0$, such that the first two equations give a slight modification of hydrodynamic thermal instability:
\ba
\sigma^2 - t^{-1}_{\rm isb} \sigma + N^2_{\rm BV}\Bigg(\frac{k_{\rm y}\delta F_{\rm Ty}}{\mathbf{k}\cdot\mathbf{\delta F}_{\rm T}}\Bigg)=0
\label{eq:type1}
\ea

While this appears to be a simple modification, this may have importance consequences for the saturation of local thermal instability in non-linear multidimensional simulation since this reduces the stabilizing impact of buoyancy (or removes effect of entropy gradient, see \citealt{2019MNRASchoudhury}). For this modified overstability, the exact magnitude of the magnetic field may not be relevant when the wave number is large along stratification or otherwise if the tension along parallel direction is negligible (although that implies large amplitude ripples in transverse direction that eventually may lead to type II). If the field aligned length scale is longer (large $k_{\rm x}$), the effect is more prominent. This is equivalent to the picture of flux tubes moving transversely under buoyancy. Unless there is a convection, there is no other linear stability problem. However, in a multidimensional atmosphere, motions in different directions are coupled and radiative cooling aids that in this 2D case.

Now let us assume that uneven non-isobaric radiative cooling generates a motion along the field (last velocity projection equation). The velocity generated along field is $\mathbf{v}\cdot\mathbf{B}_0$ in terms of which we derive the possible modes assuming $\mathbf{v}\cdot \boldsymbol{\nabla} s_0=0$. Note that by this definition, these modes strictly become type II in a formal sense. We now only use the first and last equations to express growth rate as,
\ba
\sigma (\sigma + t^{-1}_{\rm isc})= (\sigma - t^{-1}_{\rm tisb})\Bigg( \frac{(\mathbf{k}\cdot\mathbf{\delta F}_{\rm T} - k^2 \delta P_{\rm M})c^2_{\rm s} (\mathbf{k}\cdot\mathbf{B}_0)}{(\mathbf{k}\cdot\mathbf{g})(\mathbf{v}\cdot\mathbf{B}_0)} \Bigg)
\label{eq:type2}
\ea
If the term in the second bracket in RHS is positive, or in other words the projected magnetic tension along the propagation direction is larger than projected magnetic pressure gradient, an overstability occurs. In order of magnitude, the inverse timescale is $\sim \frac{k^2 v_{\rm A} \gamma H_{\rm p}}{\epsilon} \frac{\delta B}{B_0} \propto \frac{k^2 c_{\rm s} H_{\rm p}}{\epsilon \sqrt{\beta}} \frac{\delta B}{B_0} \propto  \frac{k^2 c_{\rm s} H_{\rm p}}{{\beta}^{\frac{3}{2}}} \Big[\frac{\delta B}{B_0}/\frac{\delta p}{p_0}\Big]$ assuming from the last projected momentum equation, $v_{\rm y}\sim \frac{k_{\rm y} c^2_{\rm s}}{\sigma} \frac{\delta p}{p_0} \sim v_{\rm A} \Big(\frac{c^2_{\rm s}}{v^2_{\rm A}} \frac {\delta p}{p_0}\Big) = \epsilon v_{\rm A}$ (for weak magnetic field) with $\epsilon \lesssim 1$ and $H_{\rm p}$ is the pressure scale height. In fact, in the above analysis, the difference in the magnetic forces (numerator) is $v^2_{\rm A} k_{\rm x}k_{\rm y}\Big(k_{\rm y} \frac{\delta B_{\rm x}}{B_0} - k_{\rm x}\frac{\delta B_{\rm y}}{B_0}  \Big)$. Thus $k_{\rm y}\rightarrow 0$ is stabilized/decaying (unlike type I above). These modes, if growing, must have spatial periodicity along the background magnetic field. Note that the growth may also depend on the angle between the direction of propagation of the wave and the background field. Both types of modes described above are present in our global analysis. In both global and local analysis we avoid the effect of background field curvature (not associated with the field perturbations) which, in the absence of gravity, produces type II ballooning modes. 

After understanding the driving mechanisms of idealized type I or II overstabilities, we write the following complete dispersion relation in our local analysis:
\ba
\nonumber
\sigma^5 + \frac{\sigma^4}{t_{\rm isc}} &+& \sigma^3 (k^2 c^2_{\rm ms} - i k_{\rm x} c^2_{\rm s} N_{\rm BV}^2/g -ik_{\rm x} g) \\
\nonumber
&+& \sigma^2\Big[ \frac{-k^2 c^2_{\rm s}}{t_{\rm isb}} + \frac{k^2 v^2_{\rm A}}{t_{\rm isc}} - \frac{ik_{\rm x} g}{t_{\rm isc}}\Big]\\
&+& \sigma(k^2_{\rm y} c^2_{\rm s} N_{\rm BV}^2 + k^2 k^2_{\rm y} c^2_{\rm s} v^2_{\rm A}) - \frac{k^2 k^2_{\rm y} c^2_{\rm s} v^2_{\rm A}}{t_{\rm isb}} = 0
\ea
where $c^2_{\rm ms} = c^2_{\rm s} + v^2_{\rm A}$, $v^2_{\rm A}/ c^2_{\rm s} = B^2_0 /\gamma p_0$ denote the fast magnetosonic speed and Alfv\'en wave speed relative to sound speed respectively. In the absence of gravity and associated stratification, we get a simpler form,
\ba
\nonumber
\sigma^5 + \frac{\sigma^4}{t_{\rm isc}} &+& \sigma^3 k^2 c^2_{\rm ms} 
+ \sigma^2\Big[ \frac{-k^2 c^2_{\rm s}}{t_{\rm isb}} + \frac{k^2 v^2_{\rm A}}{t_{\rm isc}}\Big]\\
&+&\sigma k^2 k^2_{\rm y} c^2_{\rm s} v^2_{\rm A} - \frac{k^2 k^2_{\rm y} c^2_{\rm s} v^2_{\rm A}}{t_{\rm isb}}=0
\label{eq: disp}
\ea
The above dispersion relation has been well known in the MHD context and in the limit of $\sigma << kc_{\rm s}$ reduces to the cubic equation only. Conventionally, the solutions of the cubic in the regime $v_{\rm A}<<c_{\rm s}$ are obtained as the classic isobaric thermal instability ($\sigma_{*} \sim t^{-1}_{\rm isb}$) aka the non-propagating condensation mode and two conjugate purely propagating Alf\'ven waves ($\sigma_{*} \sim \pm i k_{\rm y} v_{\rm A}$).

However we are probably in a regime in which $N_{\rm BV}$ is dominant compared to $\sim k v_{\rm A}$ since magnetic field is weak while buoyancy is not necessarily weak. The relevant dispersion relation is rather,
\ba
\nonumber
\sigma^5 + \frac{\sigma^4}{t_{\rm isc}} &+& \sigma^3 k^2 c^2_{\rm ms} 
+ \sigma^2\Big[ \frac{-k^2 c^2_{\rm s}}{t_{\rm isb}} + \frac{k^2 v^2_{\rm A}}{t_{\rm isc}}\Big]\\
&+&\sigma (k^2 k^2_{\rm y} c^2_{\rm s} v^2_{\rm A} + k^2_{\rm y} c^2_{\rm s} N_{\rm BV}^2)  - \frac{k^2 k^2_{\rm y} c^2_{\rm s} v^2_{\rm A}}{t_{\rm isb}}=0
\label{eq: disp2}
\ea

The slow (and fast) wave frequencies in presence of buoyancy can be quite different. To explain this, we now estimate oscillation frequency in the adiabatic case (no growth/damping due to cooling or heating). In eq \ref{eq: disp2}, if we take slow cooling regime $t_{\rm cool} \to \infty$, the conventional magnetoacoustic wave dispersion is modified as,

\ba
\nonumber
\sigma^4 + \sigma^2 k^2 c^2_{\rm ms} 
+  (k^2 k^2_{\rm y}c^2_{\rm s} v^2_{\rm A} + k^2_{\rm y} c^2_{\rm s} N_{\rm BV}^2) =0
\ea

The fast and slow waves now have the following frequencies ($\sigma = -i \omega$) -
\ba
\nonumber
\omega^2 = \frac{1}{2}k^2 c^2_{\rm ms} \pm \frac{1}{2}k^2 c^2_{\rm ms} \sqrt{1. - \frac{4(k^2_{\rm y} v^2_{\rm A} + k^2_{\rm y} N_{\rm BV}^2/k^2)}{k^2 c^2_{\rm ms}}}
\ea
Particularly for the slow mode it implies that $N_{\rm BV}$ dominates the oscillatory/propagating part if $k c_{\rm ms} >> N_{\rm BV}$. Thus, eq \ref{eq: disp2} must have a pair of fast magnetosonic waves, a pair of slow magnetosonic waves (both modified by buoyancy oscillation rates) and a growing mode. Figure \ref{fig:growthlocalgen} shows the growth rates of two slow modes from the dispersion relation eq. \ref{eq: disp2}. While uneven radiative cooling can seed parallel motions, cooling rate may not dominate the overstability for large relative $N_{\rm BV}$. We will see in our result section that the short wavelength overstability appears in global analysis above frequency $N_{\rm BV}$. This is essentially a mixed mode (motions in different directions are coupled) but has mainly type II characteristics (in effective wave numbers). Non-zero pressure perturbations trigger type II at shorter timescale. However, a type I large-scale mode, which is spiralling like in pure hydrodynamic case, also appears below $N_{\rm BV}$ in global analysis and small $k_{\rm y}$ in our tailored local analysis described in section \ref{sec:local1}.

\section{Linear eigen modes and implications}
In this section we present the results from the global mode analysis. The key characteristic to discuss in the following cases is the sustenance of long wavelength modes. First we present globally spiralling overstable buoyancy modes in the hydrodynamic case and then the possible longitudinal modes (along magnetic field) in the ideal MHD case. 

In section \ref{sec:res2}, we discuss how the modes modify in presence of a weak magnetic field in the MHD global linear analysis on a 2D plane. Magnetized plasma can form global spirals only below buoyancy oscillation frequency. Above $N_{\rm BV}$, structures of shorter wavelength form. Hence in cluster cores, both types of modes can be produced. Realistically, fragmented spiral structures or structures of azimuthal long and short wavelengths are expected to coexist in surface brightness, depending on what the background plasma $\beta$ is at the time of formation of any given azimuthal length scale. 

\begin{figure*}
    \includegraphics[width=18cm]{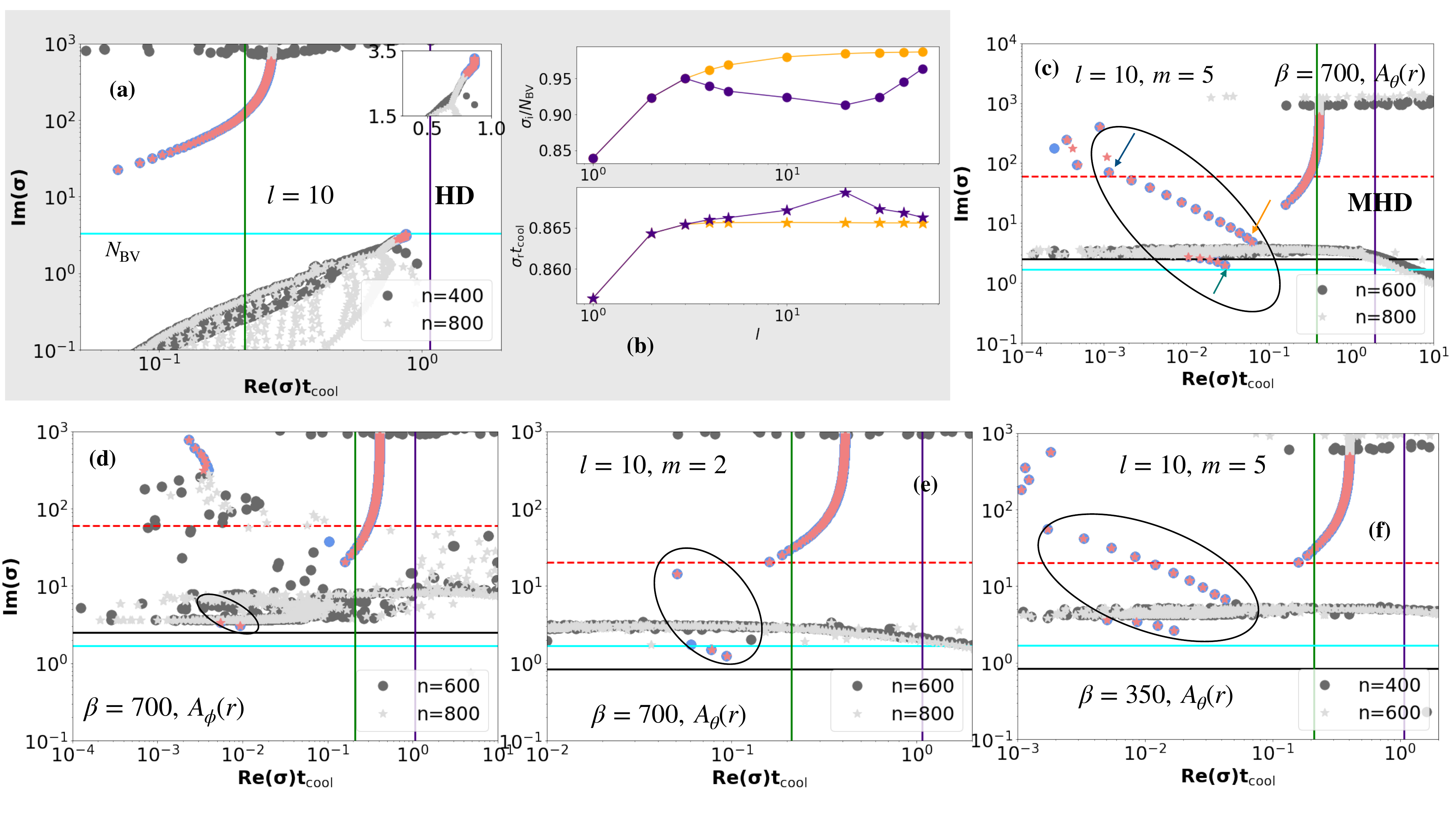}
    \centering
    \caption{Demonstration of convergence (red and blue points are converged while corresponding points which are not converged are in light gray stars and dark gray filled circles respectively) in pseudospectral method and the global growth/oscillation rates in the ICM. In the grey region, (a) shows hydrodynamic overstable buoyancy modes (cyan line corresponds to maximum $N_{\rm BV}$, the vertical lines show maximum $t^{-1}_{\rm isc}$ in green and maximum $t^{-1}_{\rm isb}$ in purple, which are characteristic buoyancy frequency, isochoric TI growth rate and isobaric TI growth rate). (b) shows the change in growth rate and frequencies for the fastest hydrodynamic mode with $l$ that sets wave number along $\hat{\theta}$. The four remaining panels show the MHD cases with overstable branches circled in black. Horizontal lines show mean of characteristic radial magnetosonic frequency (red dashed; $\sqrt{c^2_{\rm s} + v^2_{\rm A}}/dr$) and alfv\'en frequency (black solid; $v_{\rm A}/dr$) in the global box.}
    \label{fig:growthglobal}
\end{figure*}
\begin{figure*}
    \includegraphics[width=18cm]{plots/figure_2.png}
    \centering
    \caption{The eigenmodes (in global density perturbation) for hydrodynamic (grey) and MHD cases. In the former case, we show the integrated perturbations in surface brightness (assuming it is dominated by density) in the lower row. In order to calculate that at any arbitrary line-of-sight (Z$^\prime$), $\alpha$ and $\tau$ are angles by which the actual coordinate system is rotated from Z and Y axis respectively, and then the surface brightness is added along the LOS and plotted on X$^\prime$Y$^\prime$ plane. The MHD cases (for same $l$) are presented in three panels with multiple $m$ and $\beta$. Both the {\it domain} modes (inset) and the {\it reflected} mode at the boundary are shown, the former being identified as the slow compressive mode. There are no real buoyancy spiral/breathing modes at these parameters but see Figure \ref{fig:moregrowthrts} and \ref{fig:moreeigmodes2} for lower $l$.}
    \label{fig:eigmodes}
\end{figure*}
\label{sec: results}

\subsection{Large scale spiralling g-modes in hydrodynamic global analysis: perfect candidates for cold fronts}
\label{sec:res1}
In the hydrodynamic case, we explore modes with $l=10$ (related to the spherical harmonics and defined in section \ref{sec:globalsetup}) . The value only sets the coherence scale along $\hat{\theta}$ (for all scales along $\hat{\phi}$ since the hydrodynamic equations are independent of $m$ as discussed in section \ref{sec:globalsetup}) and any other value of $l$ is expected to produce similar solutions (tested in part (b) of upper left panels in Figure \ref{fig:growthglobal}; although growth rate diminishes slightly for smaller $l$). While the value of global parameter $l$ is not exactly comparable with local $k_{\theta}$, we can make a qualitative comparison of the two by matching dimension to assess a characteristic length scale, as $\frac{2 \pi}{k_{\theta}} \sim \frac{2 \pi r}{l}$, or $k_{ \theta} \sim \frac{l}{r}$ which implies at $r \sim 10$ kpc, an equivalent $k_{\theta} \sim 1$ or equivalently $\hat{\theta}$ length scale $l_{\theta} \sim 6$ kpc. For thermally overstable buoyancy modes, we expect mild dependence on local wave number in the oscillation frequency ($\sim \frac{\sqrt{k^2_{\theta} + k^2_{\phi}} N_{\rm BV}}{k}$) and none for the growth rate ($\sim t^{-1}_{\rm isb}$). We expect these long wavelength modes to have frequencies below $N_{\rm BV}$. 

In Figure \ref{fig:growthglobal}, the gray part (background of first two panels in the top row) shows the converged eigenvalues for a hydrodynamic galaxy cluster in the complex plane ({\it left}) and the fastest mode's growth and oscillation rate with $l$ ({\it right}). The former has convergence shown in red and blue colors while all the eigenvalues are shown in light and dark gray. The x-axis and y-axis in the left panel contains the absolute values and hence oppositely propagating and damping modes are also included. The modes near growth rate $\sigma t_{\rm cool}\sim 1$ (also shown in the inset) are the isobaric overstable buoyancy (see section 3.2 and APPENDIX A in \citealt{2016MNRAS_choudhury}) since the values are below the cyan line signifying maximum Brunt-V\"ais\"al\"a frequency. Both the growth rates and oscillation frequencies weakly depend on $l$ for the fastest mode (confined at small radii). The properties of the fastest mode depends on the background environment, which varies radially. For sufficiently small $l$, this variation (gradients of density, pressure , temperature, etc) must impact the mode.

The eigenvalues correspond to azimuthally small and large scale modes depending on the value of $m$. For each eigenvalue (given $l$), there are $2l+1$ eigenmodes. For small $m$, the mode is long wavelength in $\phi$ direction. In Figure \ref{fig:eigmodes}, the left upper panel (with background in gray) shows a mode with $l=10, m=2$. It is a spiral mode. The lower panel (again with gray background) shows this mode integrated along a range of line-of-sights (LOS). The mode survives in each LOS. In fact, along $\hat{z}$, the spirals become prominent (third plot in lower panel). These overstable g-modes (buoyancy oscillations) are ideal candidates that can form large scale spiralling cold fronts seen in the core of Perseus core. In practice, the large spirals will emerge in presence of a moderately large scale perturber, e.g., sub-structure passage or any stirring event like a nearly isotropic, gentle AGN feedback as is often expected in relaxed cluster cores. Since these modes are growing in density, the mass loss due to any mixing at the interface with hot medium can be replenished. However, magnetic field along cold fronts is usually believed to be preventing mixing. In the next section, we explore if presence of aligned field destroys the instability itself. 

\subsection{Effects of weak magnetization}
\label{sec:res2}
We now explore whether magnetic field supports long wavelength along $\hat{\phi}$. We take a range of plasma $\beta$ and $l$, $m$ to obtain the spectra of magnetized modes.

In MHD, both $l$ and $m$ are important parameters entering into the coupled equations. Thus, $k_{\theta} \sim \frac{l}{r}$ and $k_{\phi} \sim \frac{2\pi}{2\pi r/m} \sim \frac{m}{r}$. For $m=2,5$ at $10$ kpc, $k_{\phi} = 0.2, 0.5$. This also implies the length scales of fluctuations we pick up are $l_{\theta} \sim 6$ kpc and $l_{\phi} \sim 12$ kpc (at $k_{\phi} = 0.5$) or larger in the $r-\phi$ plane (while keeping $l=10$). Thus these parameters may lead to sufficiently long wavelength along $\hat{\phi}$, relative to that in the radial direction. We also carry out searches with smaller $l,m$ for longer wavelength modes at high $\beta$. The latter is of type I while the former should be of type II. In what follows, we describe our exploration step-by-step.

Firstly, we find that a wider range of modes are overstable when $\beta$ is a few hundreds (e.g., $\beta=350,700$), but at lower growth rates than $t^{-1}_{\rm isb}$ by a factor of $\sim 0.05-0.1$ at $l=10,m=5$ (see panels in Figure \ref{fig:growthglobal} without gray backgrounds). From our description in section \ref{sec: physset}, we know that modes with significant velocity perpendicular to the $r-\phi$ plane are not captured in this set-up. Thus transverse waves along $\hat{\theta}$, if unstable, are unavailable in this global analysis. In a 3D realistic simulation, a larger range of modes can be triggered due to this reason and the saturation properties may depend on this factor.

\begin{figure*}
    \includegraphics[width=18cm]{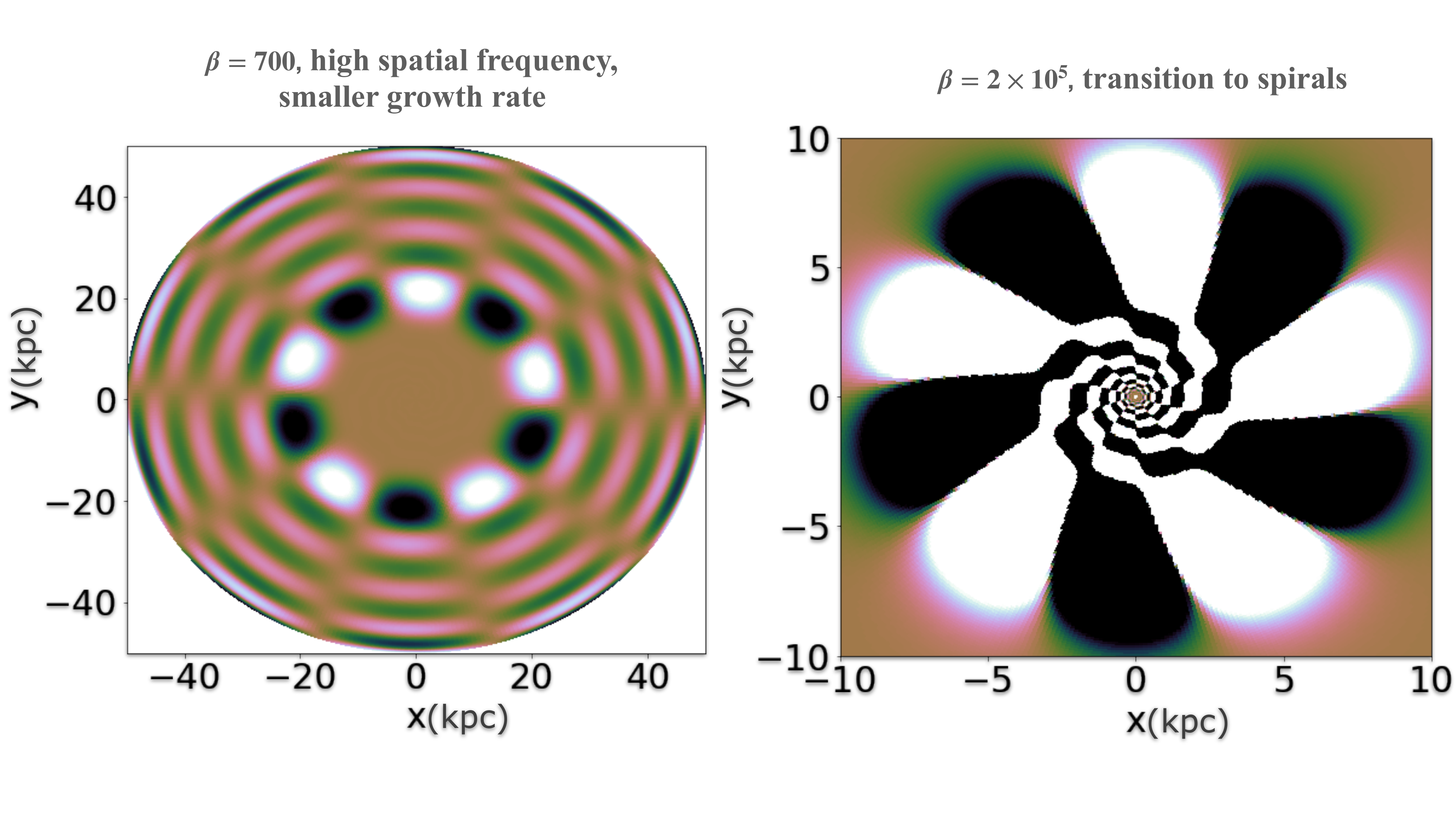}
    \centering
    \caption{In the left panel, a high frequency {\it domain} mode is shown and it is volume-filling. The frequency is marked by a blue arrow in Figure \ref{fig:growthglobal} in the upper rightmost panel. The corresponding $l=10$ as is used in the hydrodynamic case. On the right panel, the transition to spiral at extremely large $\beta$ is demonstrated and this is a buoyancy mode with higher growth rate (see Figure \ref{fig:app} for the parameter space of growth rate and oscillaion rate).}
    \label{fig:moreeigmodes}
\end{figure*}

With smaller $\beta$, the growth rate tends to decrease slightly (expected from Figure \ref{fig:growthlocalgen} as the growth reduces for stronger field). In Figure \ref{fig:eigmodes}, we show the two fastest modes for each case (indicated by the green and yellow arrows in the upper right panel in Figure \ref{fig:growthglobal}) in the upper row. There is a mode close to the outer boundary (marked by green arrow in Figure \ref{fig:growthglobal}) that is probably a reflection rather than a physical mode. Physically, the mode inside the domain is a robust overstable mode. Note that the physical location may vary for this mode (unlike the one closer to outer boundary). We define this to be {\it domain} mode in our system. In Figure \ref{fig:moreeigmodes} (left panel), we show a higher order slowly growing {\it domain} mode (marked by a blue arrow in upper right panel in Figure \ref{fig:growthglobal}). This is a volume-filling mode but short azimuthal length scales. We understand two characteristics of these new overstable modes so far: the growth rates are typically small, and it may decrease with $\beta$ (as indicated by the slow wave overstability in section \ref{sec:analyticsetup}). Further, these modes appear to be smaller in length scales than spirals (similar to type II). This can naively indicate support towards small-scale growth and hence turbulence. We carry out a more comprehensive analysis in what follows. 

If we keep increasing $\beta$, the {\it domain} modes do not disappear. However, at $\beta = 2 \times 10^5$, we find that a fraction of the hydrodynamic overstable buoyancy modes reappear (see Figure \ref{fig:app}). The latter branch is identifiable by the portion approaching $N_{\rm BV}$ (cyan horizontal line).  A branch of the overstable {\it domain} modes still persists at higher oscillation rates. If we take the fastest growing {\it domain} mode we find that the radial location is closer to the center (last panel Figure \ref{fig:app2}). This means that the physical scales of the fluctuations are smaller. Basically, the fastest growth happens at smaller scales at high $\beta>>1$. From the local analysis, the growth rate of type II (discussed after eq. \ref{eq:type2}) is $\Gamma \sim \frac{k^2 H_{\rm p} c_{\rm s} \delta_{\rm pB}}{\beta^{3/2}}$ where $\delta_{\rm pB}$ is the ratio of relative pressure fluctuation to relative magnetic field amplitude fluctuation at the radial location. Hence $\Gamma^{1/2}_{1} \beta^{3/4}_1 \delta^{1/2}_{\rm pB, 1}/\Gamma^{1/2}_{2}\beta^{3/4}_2 \delta^{1/2}_{\rm pB, 2} \propto k_1/k_2 \propto r_2/r_1$ where $r$ denotes radial location. If increasing numbers denote low to high $\beta$, the radial location for higher $\beta$ is expected to be deeper inside the core or in other words, the fastest modes have the length scale conducive to growth. The radial location for this case is indeed at $\sim 10$ times smaller radius (first and third panel of Figure \ref{fig:app2}) which matches if $\Gamma_1 \approx \Gamma_2$. But at lower radii, another possible boosting factor for growth is also the higher $N_{\rm BV}$ in the global atmosphere (see Figure \ref{fig:growthlocalgen}). This effect of $N_{\rm BV}$ vanishes when we compute the above growth rate (also from eq. \ref{eq:type2}) for {\it strictly} type II modes while in reality faster buoyancy oscillations clearly enhance the growth of the slow mode if we solve the complete dispersion relation. Physically, stronger buoyancy oscillations (high $N_{\rm BV}$) also lead to some parallel (to field) motions and hence may cause type II. Thus there is difference in growth rates in the two cases, namely, at high $\beta$ the fastest growth rate increases. As a result the type II still appears at a sufficiently large radius visible within our domain. On the other hand, the reappearing buoyancy mode transits to the spiralling hydrodynamic eigenmode (right panel of Figure \ref{fig:moreeigmodes}). The latter is a test for the transition to hydrodynamic case. 

\begin{figure*}
    \includegraphics[width=18cm]{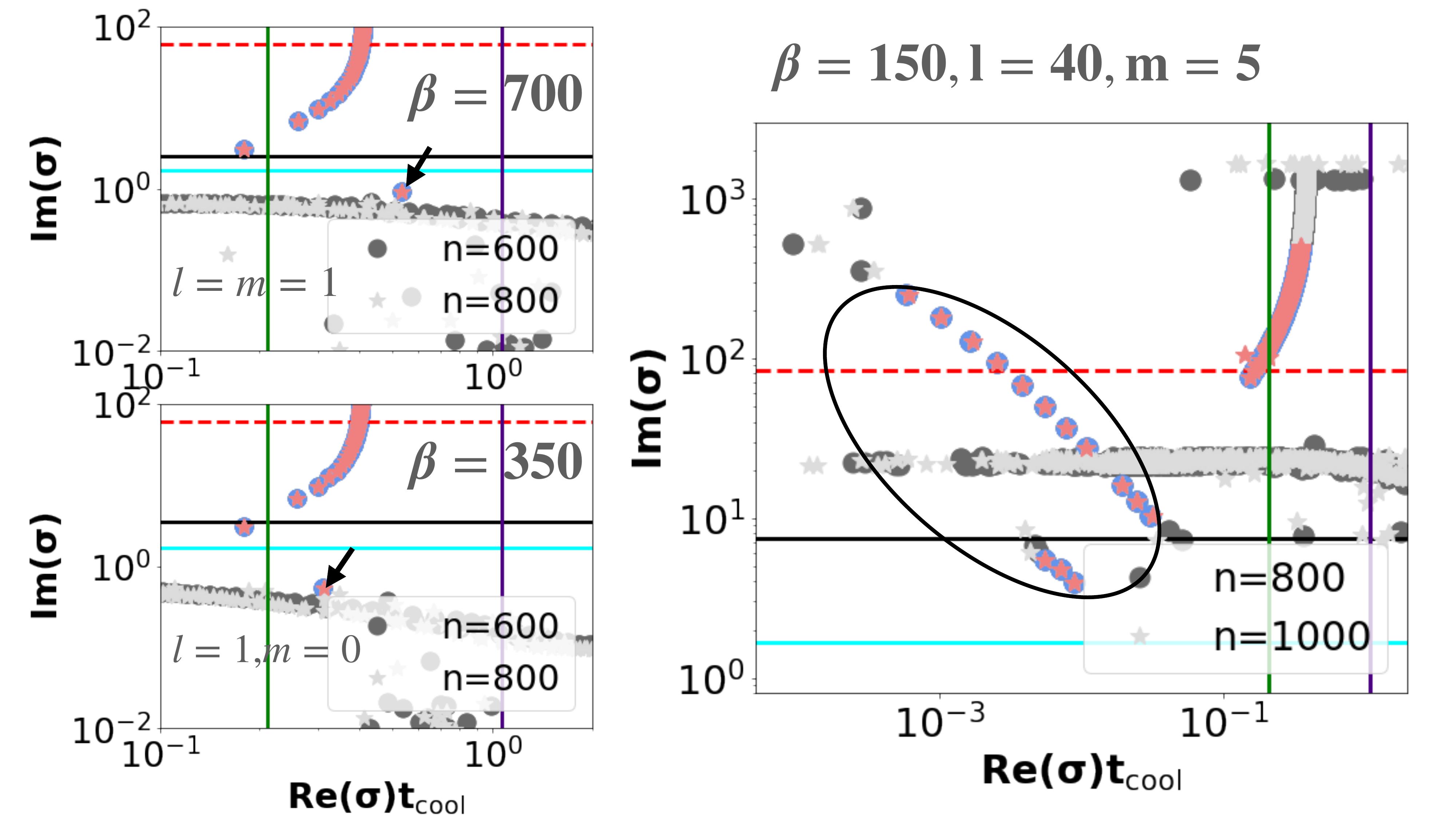}
    \centering
    \caption{The growth (damping) rates (converged points in red and blue while corresponding points which are not converged are in light gray star and dark gray filled circles respectively) and oscillation rates of more global overstable modes for lower values of $l$ (left) and large values of $l$ (right). The former cases show emergence of azimuthal long wavelength spiral/breathing modes. At large $l$, no such modes are easily found in our exploration. }
    \label{fig:moregrowthrts}
\end{figure*}

\begin{figure*}
    \includegraphics[width=18cm]{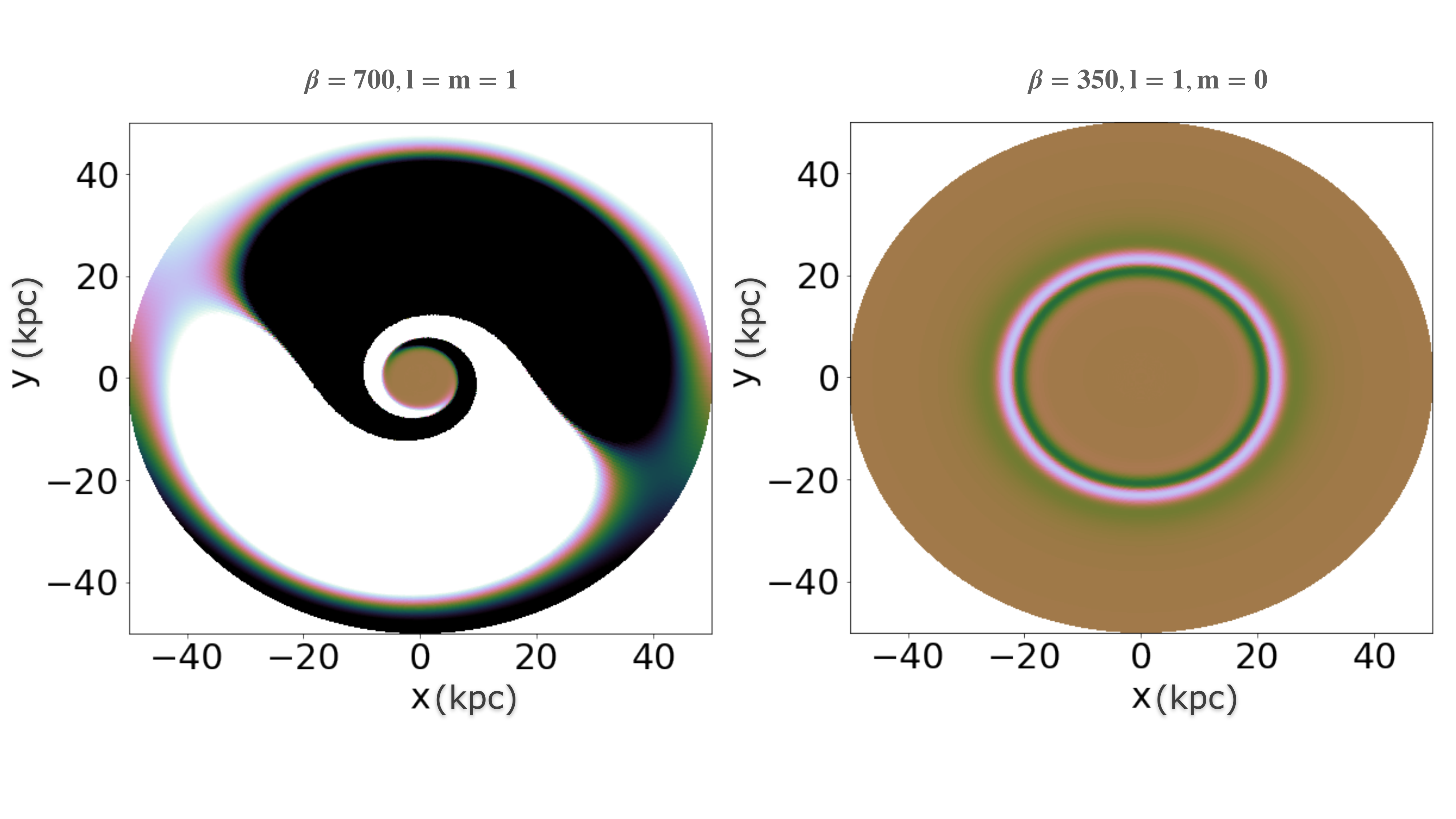}
    \centering
    \caption{Two large scale (azimuthal) modes obtained for weakly magnetized cases at frequencies below $N_{\rm BV}$, the characteristic buoyancy oscillation frequency. The left panel shows large scale spiral while the right panel is a symmetric-spherical mode. Availability of these modes at large growth rates in the weakly magnetized cases confirm that spiralling is a persistent characteristic for global buoyancy modes irrespective of plasma $\beta$. While these are rarer in presence of magnetic field, these are expected to be robust growing features in weakly magnetized ICM core at sufficiently long timescales. }
    \label{fig:moreeigmodes2}
\end{figure*}

We further use smaller values of $l$ and $m$ at $\beta=700, 350$ and a large $l$ at $\beta=150$ in Figure \ref{fig:moregrowthrts}. The purpose of the former exploration is to find if global spirals exist at the largest scales in weakly magnetized medium. For the first two cases, we find such azimuthal structures (Figure \ref{fig:moreeigmodes2}) while for the third case we mostly find all high-frequency unstable slow modes. In section \ref{sec:analyticsetup}, we discuss the two broad types of modes and the above exploration clearly extracts type I at large $\beta$. For a combination of large $[l,m,\beta]$, retrieving the isobaric mode is difficult. We cannot conclusively determine strongly magnetized cases since our background equilibrium is hydrodynamic (gravity and pressure gradient are the strongest forces) and the effect of magnetic fields may interfere with the consistency of background equilibrium. All we can conclude is that at a few hundred $\beta$, the spirals/symmetric-spherical modes (type I) may exist along with the type II. We present examples in Figures \ref{fig:moregrowthrts} (top and bottom in the left panel) and Figure \ref{fig:moreeigmodes2} of the largest scales obtained for the aforementioned $l,m$ combinations, noting that these have higher growth rates (approaching isobaric rates).

In order to understand all the mentioned characteristics of {\it domain} modes that we see in global cluster atmosphere, we investigate the local properties of waves and instabilities now. This helps us to confirm the nature of the overstability and its local dependence on scales (wave number in the local dispersion relation). While local and global behavior may not match precisely (due to absence and presence of gradients of background atmosphere), we claim that the identification of the relevant mode is valid. Further, we will assess from this analysis the effect of the overstabilities on the disruption of the spiral. 

\begin{figure*}
    \includegraphics[width=18cm]{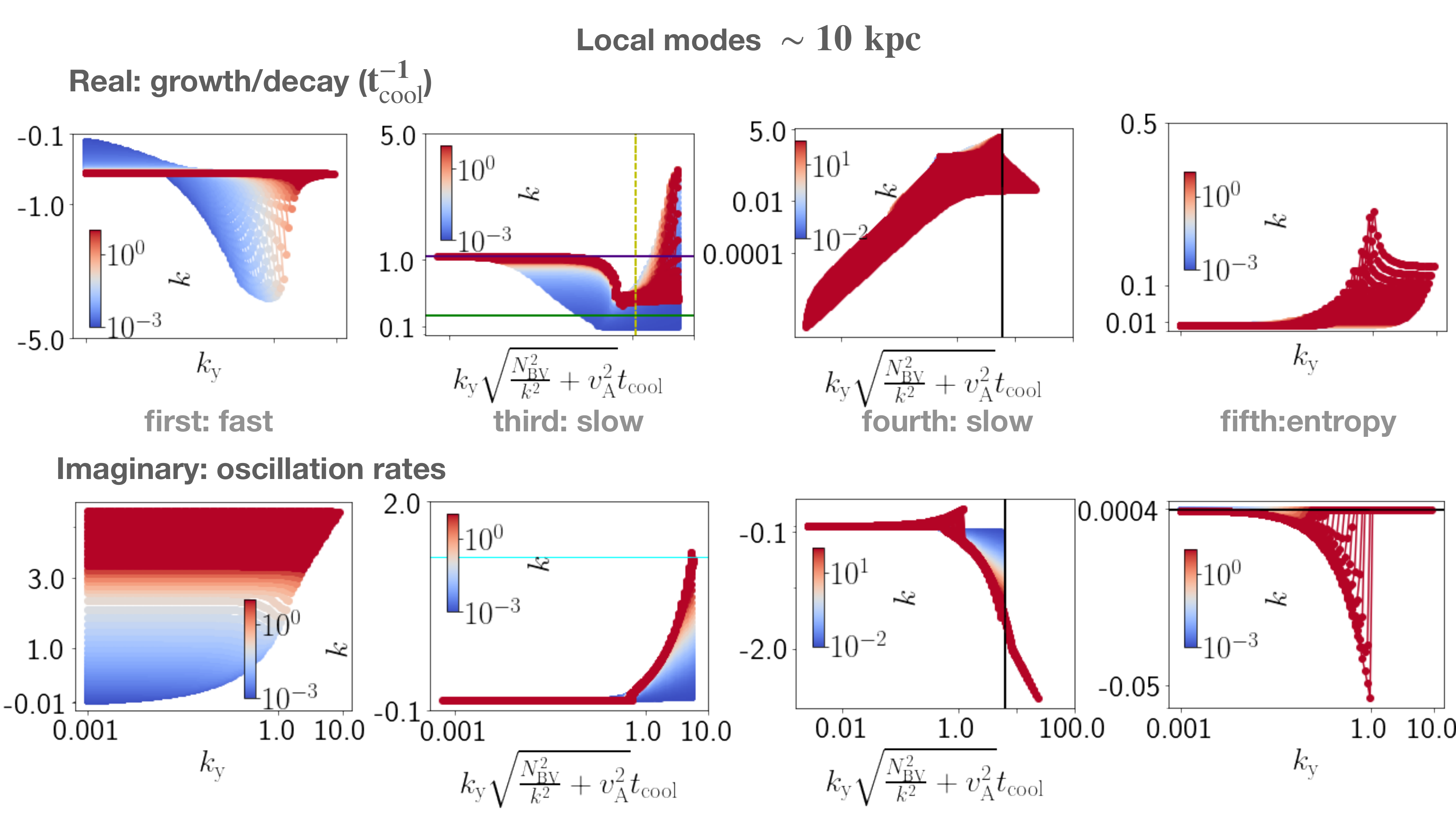}
    \centering
    \caption{The analytic local modes of a simple model of 2D plane defined by the direction of gravity and magnetic field. The physical parameters of density, pressure, cooling time, etc are extracted for the location of the overstable {\it domain} mode in the global analysis with $\beta=700, l=10, m=5$. Five modes are present in this system including two pairs of fast and slow compressive modes and a purely growing mode. The upper panel consists of the four growth rates and the lower panel shows the oscillation rates of the modes (excluding one of the fast magnetosonic modes which is identical to the first but oppositely directed). The second and the third panels in both rows show the slow compressive overstable modes. The yellow dashed line in second panel and top row represents $k_{\rm y} \sim \sqrt{k^2_{\phi} + k^2_{\theta}} = 1.12$ (see section \ref{sec:res2}). The purple and green horizontal lines show local $t^{-1}_{\rm isb}$ and $t^{-1}_{\rm isc}$ which implies the global growth rate for type II is smaller than the local estimate (in red) by a factor of $\gtrsim 2$ and expected to be well below $t^{-1}_{\rm isb}$. If type I modes appear, these will always dominate the growth for a large range of $k_{\rm y}$. The black vertical lines in third column denote $k_{\rm y} \sqrt{\frac{N^2_{\rm BV}}{k^2} + v^2_{\rm A}}t_{\rm cool}=2 \pi$.}
    \label{fig:localmodes}
\end{figure*}
\begin{figure*}
    \includegraphics[width=18cm]{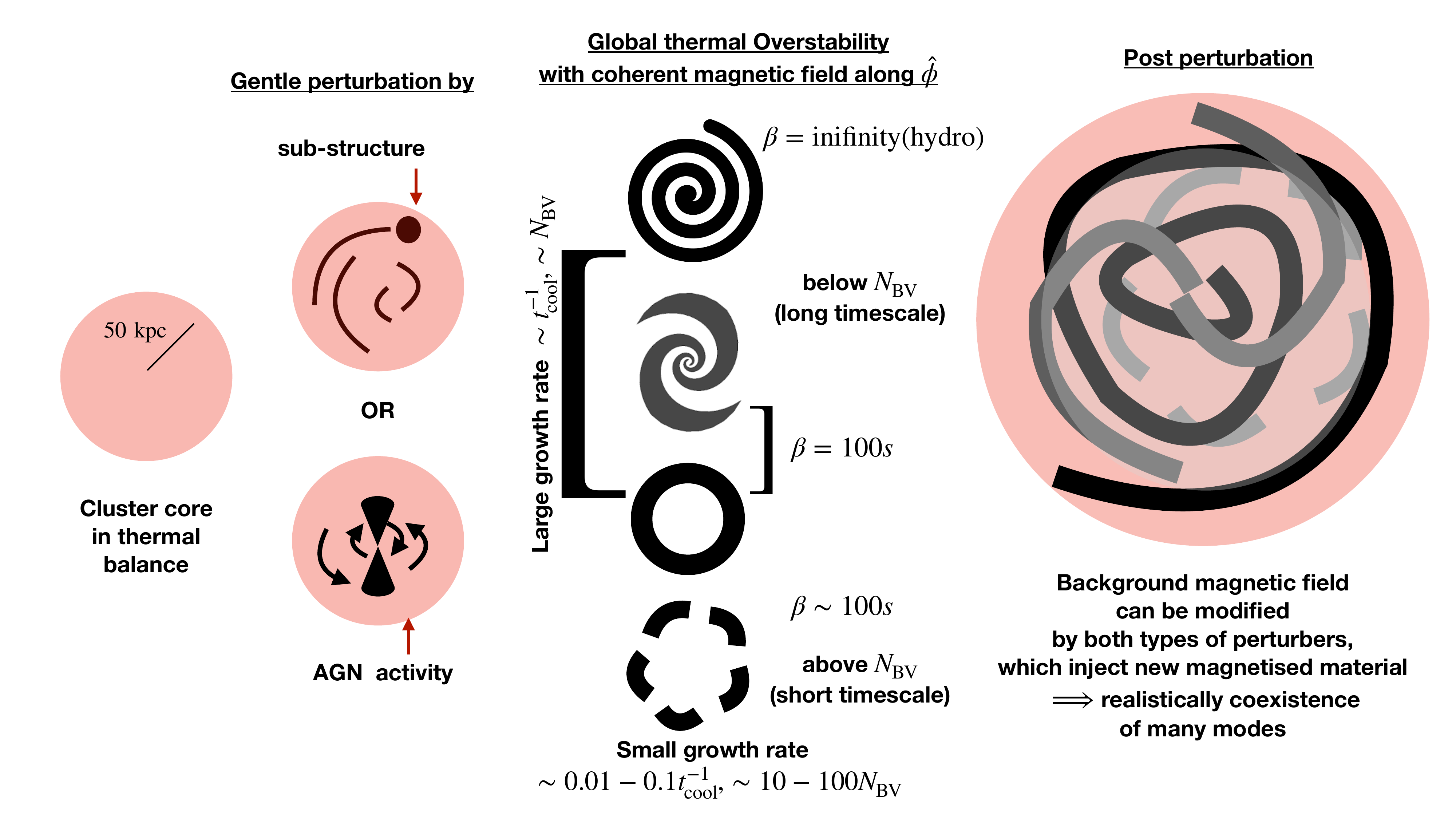}
    \centering
    \caption{A cartoon demonstrating the physically realisable scenario of the modes in relaxed galaxy cluster cores (e.g., Perseus). Perturbations at sufficiently large spatial scales from either sub-structure passage or gentle AGN feedback can generate multiple modes of a range of frequencies and wavelengths in the core. Out of these modes, the buoyancy modes are spiral and thermally unstable in the hyrodynamic limit ($\beta = \infty$). At $\beta=$ few hundreds, spirals and spherically symmetric modes may form as well but these are rarer and only below the characteristic buoyancy frequency. Above that frequency, all modes are fast propagating, overstable slow compressive modes. The magnetic field (or $\beta$) in the background medium can change over time due to injection of fields by the AGN or the satellite gaseous halo. Moreover, modeling the ICM core with a coherent magnetic field in $\hat{\phi}$ is also unrealistic. Consequently, the core is infested with many modes of varying characteristics in reality and not just one of these modes. The robust prediction of all these modes is the filamentary nature of these with lengths aligned along a local coherent magnetic field.  }
    \label{fig:cartoon}
\end{figure*}
\subsubsection{Interpretation of the global linear overstable modes using local analysis}
\label{sec:local1}

In order to interpret our {\it domain} modes, we take the approximate location of the fastest {\it domain} mode in the case with $\beta=700$ from global analysis (Figure \ref{fig:growthglobal}) which is $\sim 10$ kpc. We retrieve the physical parameters (e.g., $t_{\rm cool}, N_{\rm BV}, v_{\rm A}, c_{\rm s}$) at the radial location, plug those in the dispersion relation described in section \ref{sec:analyticsetup} (eq. \ref{eq: disp2}). Thus we are able to see the consequence of coupling between magnetoacoustic waves and buoyancy in presence of thermal instability. Figure \ref{fig:localmodes} shows the growth rate (upper panel) and frequencies of oscillations (lower panel). The modes are ordered by the highest to lowest frequencies from left to right (we do not show the second fast mode, identical but oppositely directed to the first). The y-axis is normalized as Figure \ref{fig:growthglobal} and the $k_{\rm y} \in [0,k]$. The slow modes show growth (middle panels) and differ in characteristics at sufficiently large scales (small $k_{\rm y}$) depending on the direction of the wave along (positive frequencies) or against (negative frequencies) the background field. In the second slow mode, we have extended the range of $k_{\rm y}$ to test the behavior at larger values. The other mode in the slow pair behaves similarly at high $k_{\rm y}$. The growth rate is the highest at $k_{\rm y} \sqrt{\frac{N^2_{\rm BV}}{k^2} + v^2_{\rm A}} t_{\rm cool} = 2\pi$ or in other words at length scale covered by slow mode in a cooling time. This is intuitively clear as the mode can grow only if it doesn't propagate fast. Using the local dispersion, we tested the behavior of these modes (at the same radial location of the global mode) by increasing $v_{\rm A}$ by several factors while keeping other parameters constant. We note that the slow mode growth gradually must disappear for stronger magnetic field. The peak growth shifts to smaller $k$ and lower growth rate by a few factors at $\sim 40 v_{\rm A}$, and at $\sim 1000 v_{\rm A}$. The fifth mode is the only dominant mode with small growth rate at large $v_{\rm A}$. Note that we cannot test such strongly magnetized regime in global context since we need to modify the background global equilibrium to test low $\beta$ regimes. 

In order to interpret the global modes, the local analysis provides our guiding principles. The high-frequency overstability that we find in the global analysis ({\it domain} modes), must be the overstable slow modes that oscillate at frequencies $\gtrsim N_{\rm BV}$ (as is evident from the local dispersion relation and global analysis). In Figure \ref{fig:localmodes}, there is evidence that this compressive overstability has a strong wavenumber dependence. In the third column we show a larger range for $k_{\rm y}$. The oscillation frequencies in second and third are identical except the direction of the wave. The growth rate varies for the slow modes in opposite directions (see discussion in section \ref{sec:analyticsetup} about wave number dependence in local analysis). In the global analysis, overstable modes arise in cases with several combinations of $[l,m]$. A large-scale spiral only occurs when the oscillation frequency is below the maximum $N_{\rm BV}$ (and also below the local values of $N_{\rm BV}$). To be precise, above $N_{\rm BV}$ the {\it domain} modes are overstable slow modes (type II) and below $N_{\rm BV}$ these are quite similar to isobaric buoyancy modes (type I). From local and global analysis, it is clear that large-scale buoyancy modes may not be abundant in magnetized medium. Since the ICM is weakly magnetized, we claim that type I azimuthal/spiral modes form over a cooling timescale as expected, along with shorter azimuthal wavelength type II modes depending on the spatial distribution of $\beta$. A global picture is conceptualized with the summary image in Figure \ref{fig:cartoon}. 

 The overstable slow modes ({\it domain} modes) are growing only in a narrow range of scales locally (the length scale a slow mode crosses in $t_{\rm cool}$). Further, the peak of type II is prominent if $\beta$ is high (as tested in the local analysis; but type I also reappears at large growth rates as shown in Figure \ref{fig:app}). Figure \ref{fig:moregrowthrts} (right panel) also shows $\beta = 150$ case with all overstable slow modes above $N_{\rm BV}$ and one reflected mode at the lowest frequency. This implies at small radial scales and large azimuthal scales, the slow mode grows slowly. Thus regions, with ordered strong and weak background magnetic fields side by side, are predicted to develop fast growing spirals and slowly growing smaller azimuthal scales. On the other hand, at same $\beta$ (high), depending on frequencies and length scales, both spirals and smaller azimuthal scales can coexist. 
 
\section{Discussions and Conclusions}
\label{sec:discnconc}
In this work, we propose that a wide range (in azimuthal scales, growth rates, radial location, etc) of thermally unstable modes in unmagnetized and weakly magnetized plasma can simultaneously explain the presence of large scale spiral cold fronts (type I) and small scale sub-structures (type II) in and around these. This idea is closely aligned with asteroseismology (e.g., \citealt{1994ARA&A_brown}) in which an inverted problem is considered from solar/stellar normal modes to estimate the physical conditions in the core of such objects. Multiphase and possibly filamentary plasma is present in the solar atmosphere analogous to the cluster cores (for example, discussed in \citealt{frontier_choudhury}). Here we are assessing if eigen modes may exist in the ICM at sufficiently large scales. Perseus is an example of galaxy cluster in which such a mode analysis can be explored using future X-ray mission at high spatial resolution like AXIS (\citealt{2024Univ_russell}). A tentative picture of a relaxed cluster core filled up with eigen modes is given in Figure \ref{fig:cartoon}. Further, these modes are possibly relevant for the circumgalactic medium (CGM), which is basically scaled down in mass, size, and other properties compared to the ICM (e.g., Figure 4 and Figure 8 in \citealt{2019MNRAS_choudhury}). Magnetic field strengths in the CGM is unknown except some upper limits ($\lesssim \mu {\rm G}$) from Faraday Rotation measures (e.g., \citealt{2020MNRAS_lan}) which translates to high plasma $\beta$. 

We find that large scale spiral modes and breathing modes are expected in the galaxy cluster core in unmagnetized and weakly magnetized cases, although in latter such modes are rarer. At timescales longer than inverse Brunt V\"as\"al\"a frequency ($N^{-1}_{\rm Bv}$), such modes form in weakly magnetized plasma (high $\beta$). These are also thermally unstable within comparable timescales ($t_{\rm cool}^{-1} \sim N_{\rm BV}$). At higher frequencies than $N_{\rm Bv}$, we find that the slow wave is overstable in a local approximation. In the global atmosphere, depending on the length scales and frequencies, both slow wave and buoyancy wave overstabilities may coexist. 

There is a large uncertainty about the morphology of magnetic field in the cluster core. A combination of multiple techniques like Faraday Rotation (e.g., see \citealt{2021NatAs_digennaro} for a discussion on how and when cluster magnetic field reached the current amplitude of a few $\mu G$) and Synchrotron intensity gradient (e.g., \citealt{2024NatCo_hu} for a recent discussion) have suggested that coherent magnetic field may span a sufficiently large range of scales ($5-500$ kpc) compared to the cluster core length scale ($100-200$ kpc). In fact, the above work supports the case for elongated magnetic field lines along the direction of merger axis for merging clusters; this hints that any global perturber can produce coherent field in a given direction. Moreover, a kinetic jet from the central AGN activity may also produce misaligned density and temperature gradients in the diffuse medium ($\boldsymbol{\nabla}p \times \boldsymbol{\nabla} \rho$) and hence a coherent magnetic field in the plane perpendicular to jet axis. Radio images of galactic atmospheres often reveal relatively large scale coherent filamentary magnetic field produced by possibly jet activities or unknown processes (e. g., discussed in \citealt{2021NatAs_kale}, \citealt{2022ApJ_yusuf-zadeh} for Milky Way halo, \citealt{2022ApJ_rudnick} for jet-cluster interaction, \citealt{2022ApJ_rajpurohit} for filaments in merging clusters and so on). Recently, \citealt{2024ApJ_omoruyi} show closely aligned multiphase gas in merging galaxies in a cluster environment in which diffuse radio, H$\alpha$, and molecular phases have been revealed along a $25$ kpc arc. This X-ray deficient region is different from cold front that we envisage in this work, but such a region can be an end-state of saturated instabilities that we find in this work. It is essential to assess the non-linear stage of our instabilities to conclude on the latter scenario. 

In a conservative sense, there is no concrete evidence of fully azimuthal field that we use in the ICM core. Thus we use a simple, idealized model for the field morphology ($\hat{\phi}$) which may not be the most realistic consideration for a global atmosphere. The purpose of the field is to demonstrate that if there is a coherent magnetic field along the azimuthal direction, thermally unstable spirals or arcs are easily expected. Realistically, smaller scale field structures (as opposed to a completely coherent field along $\hat{\phi}$) can be more common and lead to the development of type II modes at shorter timescale and type I at longer timescale. The first one, being oscillatory, will produce deformed type I growing modes. Our upcoming ideal MHD simulations can robustly confirm if azimuthal long wavelength modes are sustained in the core under such circumstances. 

We can speculate that instability can accumulate mass at super-Alfv\'enic characteristic velocity and that itself cause lowering of $\beta$ and saturating the instabilities. There will also be a bulk outward propagation speed of the individual global spiral modes if formed (comparable to its density growth speed) but that should cause advection of magnetic field outward instead of accumulation of field lines. However, if the medium is infested with many growing global features propagating with a range of directions/speeds, there can be highly amplified magnetic field created in the narrow compressed regions produced. Local TI may also build up magnetic field by condensation. These should probably appear in small and large scale radio filaments (e.g., \citealt{2022ApJ_yusef-zadeh}) in the clusters. 

The other missing aspect of the global linear theory is the absence of the perpendicular plane (3D) and the fifth slowly growing mode over large range of wave numbers along the background field in our local analysis (see last panel in Figure \ref{fig:localmodes}). If plasma $\beta$ reduces locally, this may grow faster. Since this grows at distinctly different $\beta$, this may not disrupt any pre-existing spiral structure. However, in order to assess this effect, a simulation of the cluster core is ideal (to be pursued in an upcoming manuscript but Appendix \ref{sec:app3d} shows an example of preliminary ideal MHD simulations).

The main conclusions from this paper are:
\begin{itemize}
    \item {\bf Spirals in hydrodynamic cluster - cold fronts: }In the hydrodynamic cluster, thermally unstable isobaric buoyancy modes produce perfect spirals of varying azimuthal length scales. Theoretically, the modes are modelled as spherical harmonics with radial dependence characterized on a Chebyshev basis. A large range of $l,m$ associated with spherical harmonics gives rise to spirals in any given radial eigenmode. These spirals, if formed, must appear in high resolution X-ray imaging of cluster cores as ``cold fronts''. 
    \item {\bf Spirals and shorter azimuthal modes in ideal MHD: } In a weakly magnetized medium, there are two modes that may grow in density. We study an idealized case in which the magnetic field is perpendicular to the gravity and is along $\hat{\phi}$. Note that there is no observational evidence of magnetic field morphology to be globally along $\hat{\phi}$ for any cluster. Such fields are conceivable in a plane perpendicular to AGN jet axis due to misaligned density and pressure gradients. Even sub-structure passage may form large-scale coherent fields perpendicular to the direction of motion. The latter may cause dragging of fields via stripping of interstellar (ISM) or circumgalactic medium (CGM). We find that large scale spirals may still form similar to buoyancy oscillations in hydrodynamics. In addition, an overstable slow wave, which propagates faster than buoyancy modes, can also grow at rates $\sim 10$ times smaller than the spirals. Although the latter is predicted to be prevalent at wide range of length scales, the growth is insignificant at most of these wavelengths. Thus slow modes cannot probably destroy a pre-existing spiral or a spiral that forms in long timescale (without or with a moderate magnetic field). 
    \item {\bf Physical reason for two modes in magnetized medium: } In section \ref{sec:analyticsetup}, we attempt to reconcile the results with previous literature on instabilities in presence of magnetized buoyancy. The spiral modes that appear in weakly magnetized medium (type I) is similar to buoyancy waves except that the oscillations happen as a flux tube in response to buoyant perturbations of fluid. The smaller (azimuthal) scale modes (type II) appear due to inhomogeneities along the field (perpendicular to gravity) that generate compression/rarefaction in that direction. In principle, even an arbitrarily small velocity, thus produced along field, can trigger weak growth for sufficiently small-scale modes ($\mathbf{k}\cdot\mathbf{B_0}$) and unbalanced tension and pressure forces. Stronger gravitational force along the propagation of wave can suppress growth (eq. \ref{eq:type2}). In reality, each type of mode will trigger the other. Whether type I or II dominates depends on timescale and source of perturbation (Figure \ref{fig:cartoon}).
    \item {\bf Future direction: }The key question, {\it can type II modes (described in previous conclusion point) destroy type I?} is addressed with the following arguments:
    \begin{itemize}
        \item type II propagates fast (global result) and grows weakly,
        \item type I is isobaric, propagates over a longer timescale (also confined spatially as g-modes) and grows faster,
    \end{itemize}
    hence the former cannot destroy the latter. We explore in an upcoming work with 3D MHD simulations (e.g., Appendix \ref{sec:app3d}), if this prediction based on 2D model holds in the non-linear 3D model.
\end{itemize}

%
%

\section*{Acknowledgements}
The authors are grateful to the anonymous reviewer for a thorough review. PPC thanks Paul Nulsen and John ZuHone for helpful discussions on cold fronts. PPC further acknowledges the plasma physics group at University of Oxford, and a helpful conference on intracluster medium, ``7th ICM theory and computation'' at University of Michigan, Ann Arbor in June 2024. 

\section*{Data Availability}
The data underlying this article will be shared on reasonable request to the corresponding author.
%




\bibliographystyle{mnras}
\bibliography{bibtex} 

%
%
%
\appendix
\section{Limiting to hydrodynamic case}
\label{sec:app}
\begin{figure}
    \includegraphics[width=9cm]{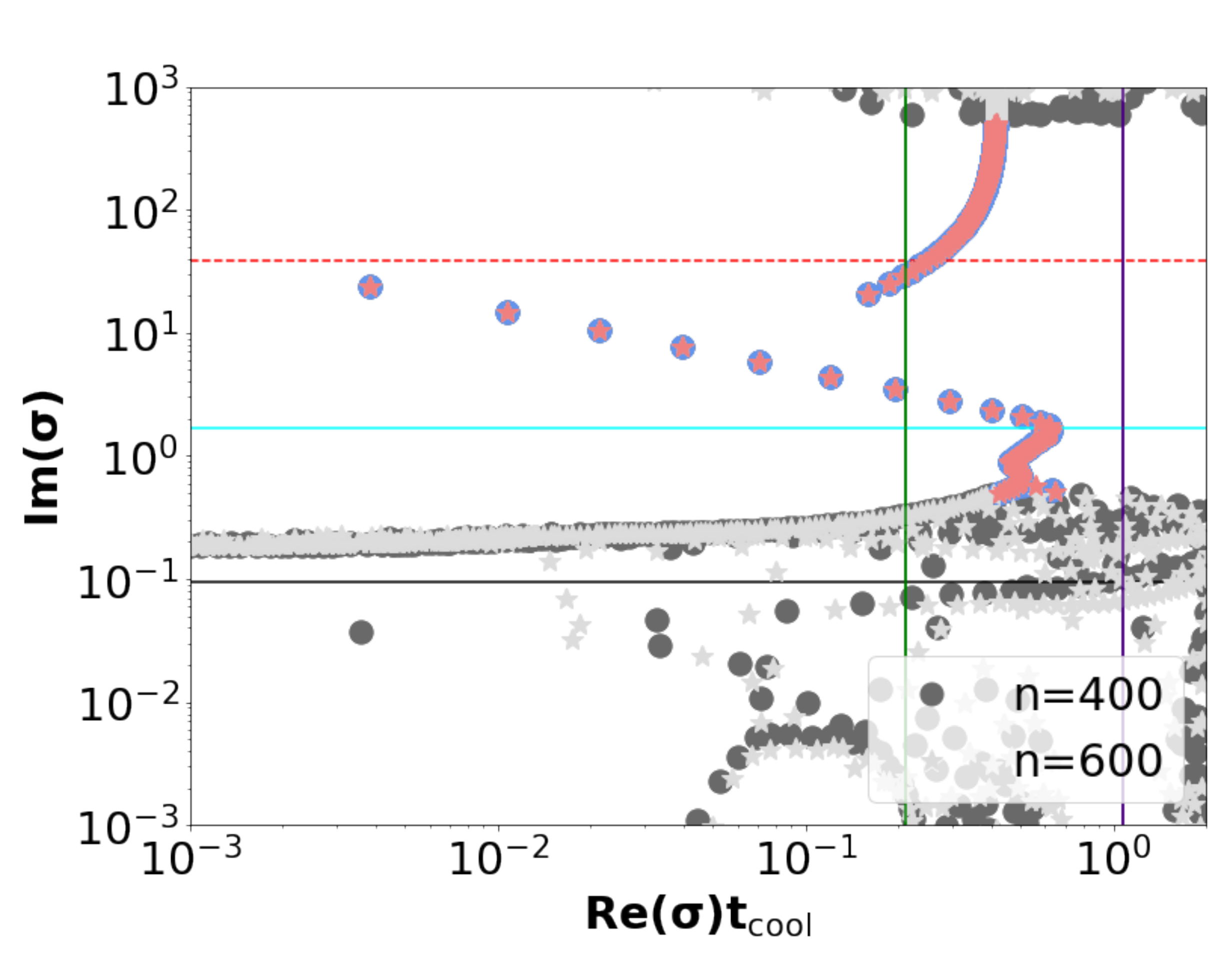}
    \centering
    \caption{Growth rates and oscillation rates for a case with $\beta=2 \times 10^5$ which shows that the branch of solution that converges shifts towards the isobaric growth rates (purple vertical line) as expected from previous literature. These are overstable buoyancy modes as evident from the prevalence of points below the horizondal cyan line denoting $N_{\rm BV}$.}
    \label{fig:app}
\end{figure}
\begin{figure}
    \includegraphics[width=8cm]{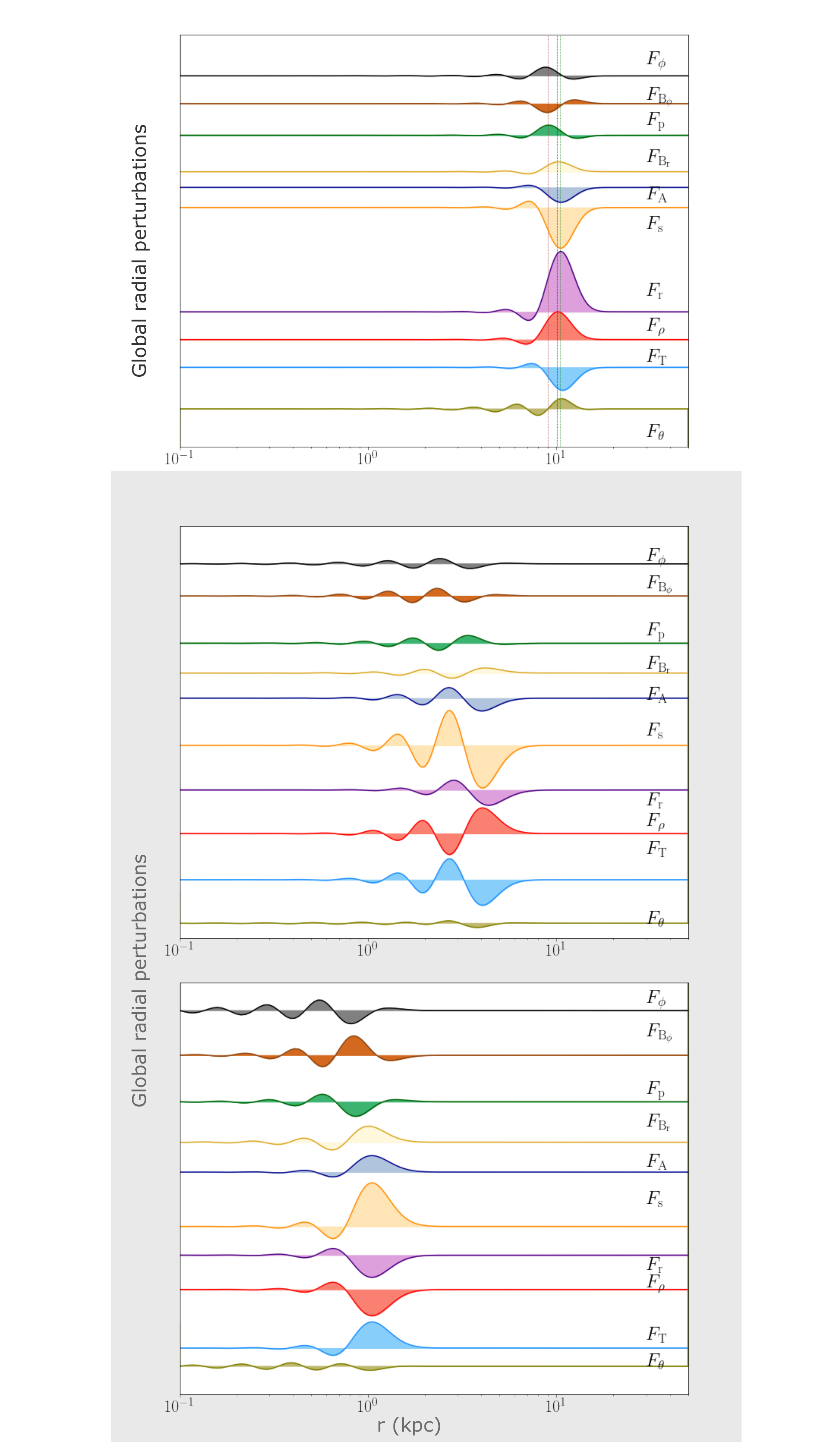}
    \centering
    \caption{The upper panel shows the fastest global eigen mode in 1D in the fiducial magnetized case of $\beta=700$. The middle and lowest panels show the overstable buoyancy and acoustic modes for $\beta=2\times 10^5$. Each color represents one variable of the perturbations and the subscripts include the name of the variables e.g., $\rho, p, T$ etc and $F_{\rm r}, F_{\theta}, F_{\phi}$ are the velocity perturbations. In all the panels $10 F_{\theta} <<F_{\rm r}$ as required by our method. }
    \label{fig:app2}
\end{figure}

We tested the hydrodynamic limit of zero magnetic field using our pseudospectral code. We find that even an infinitesimal field triggers acoustic overstability above oscillation rate $N_{\rm BV}$ (cyan line in Figure \ref{fig:app}). In this case, overstable buoyancy oscillations appear below $N_{\rm BV}$ at growth rates closer to $t_{\rm isb}$ which also appear in pure hydrodynamic case. When the magnetic tension term, proportional to the wave number of a mode, is relevant in the dynamics (e.g., magnetorotational instability), even for weak magnetic field, the wavenumber of small-scale modes becomes extremely large. Figure \ref{fig:app2} shows the one-dimensional radial eigen modes for the fiducial magnetized case and the case for hydrodynamic limit (marked in gray). In the latter case, the overstable buoyancy mode (middle panel) and slow acoustic mode (lowest panel) distinctly emerge. 

\section{A view of 3D MHD simulations}
\label{sec:app3d}
\begin{figure}
    \includegraphics[width=4cm]{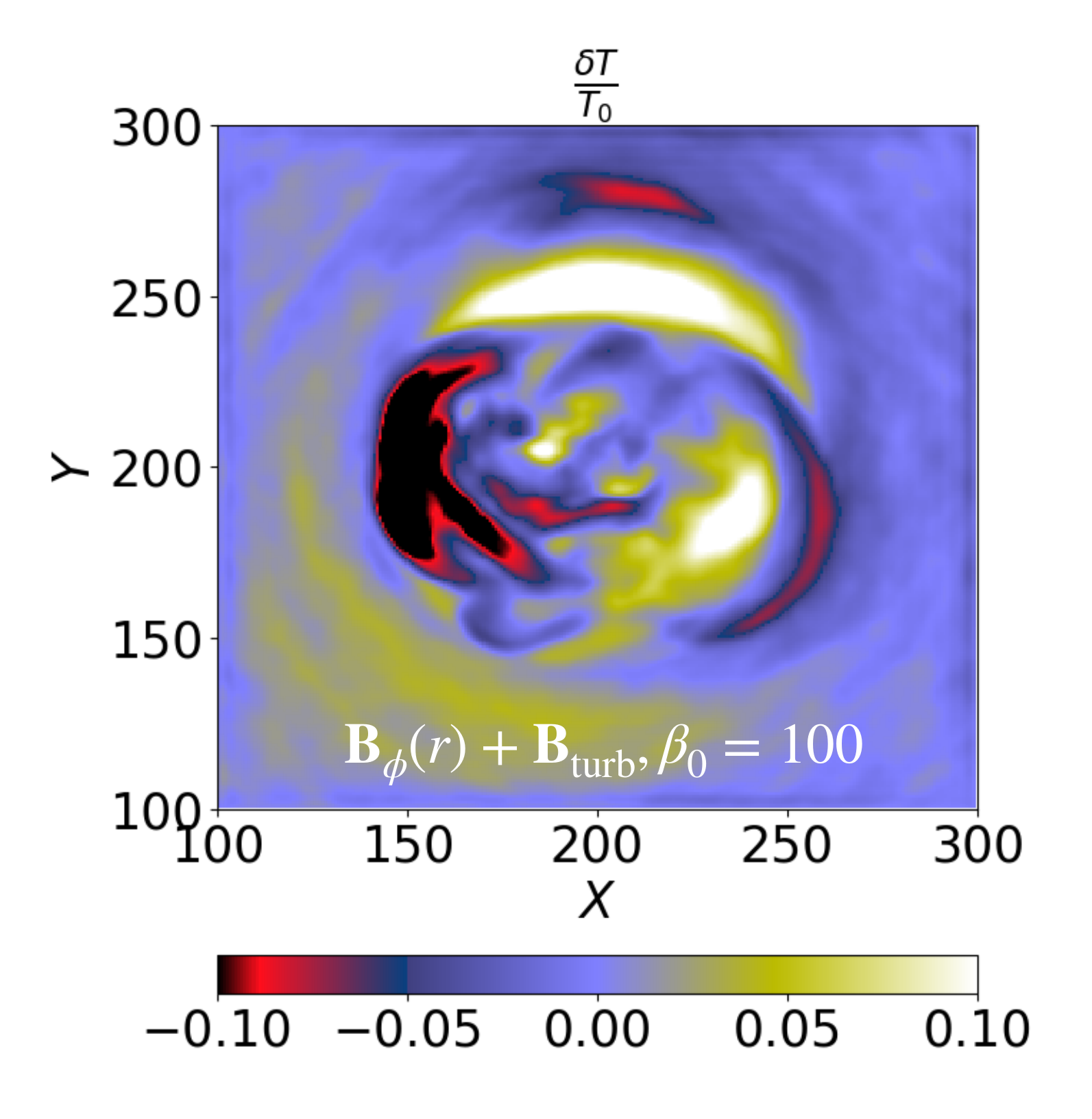}
    \centering
    \caption{The line-of-sight integrated temperature contrast in a slice of a 3D MHD simulation performed in PLUTO (a conservative hydrodynamic code with constrained transport scheme to maintain divergence-free condition for magnetic field). The magnetic field is similar to the analysis done in this paper but with additional random perturbations. The core is multiphase but closer to isobaric condition with large-scale dense features (black/red). The natural expectation, from what we understand in the current paper for isobaric state, is that large-scale azimuthal modes must easily form. }
    \label{fig:app3}
\end{figure}

In this section, we include a snapshot of one of our ongoing MHD simulations that hint at the formation of large scale azimuthal features in the cluster core. It is a $200$ kpc box with the cluster in the centre as described in \citealt{2022MNRAS_choudhury}. In addition, we include a magnetic field along $\hat{\phi}$ along with a small random perturbation in magnetic field in all directions. Our simulations demonstrate that presence of a wide range of length scales in overstability in the presence of stratification, local radiative cooling, and weak magnetic field is possible even in non-linear model.

%
%
%
%
\bsp	
\label{lastpage}
\end{document}